\documentclass[useAMS,usenatbib]{mn2e}
\usepackage{aas_macros}
\usepackage{graphicx}
\usepackage{lineno}
\usepackage{amssymb}
\usepackage[colorlinks=true,citecolor=blue]{hyperref}
\usepackage{multicol}

\newcommand{\pks}{PKS\,0625$-$354}
\newcommand{\lat}{\textit{Fermi}-LAT}

\title[H.E.S.S. discovery of very high energy $\gamma$-ray emission from  PKS\,0625-354]{H.E.S.S. discovery of very high energy $\gamma$-ray emission from PKS\,0625-354}

\newcommand{\inst}[1]{$^{#1}$}

\author[H.E.S.S. Collaboration]{\normalsize H.E.S.S. Collaboration,
H.~Abdalla\,\inst{1},
A.~Abramowski\,\inst{2},
F.~Aharonian\,\inst{3,4,5},
F.~Ait Benkhali\,\inst{3},
A.G.~Akhperjanian\protect\footnotemark[2]\,\inst{6,5}, 
\newauthor\normalsize
T.~Andersson\,\inst{10},
E.O.~Ang\"uner\,\inst{7},
M.~Arrieta\,\inst{15}
% ...
\\
Affiliations can be found at the end of the article.
}

\newcommand{\affiliations}{
\vspace{5mm}
\it\small
\noindent
\inst{1}Centre for Space Research, North-West University, Potchefstroom 2520, South Africa\\
\inst{2}Universit\"at Hamburg, Institut f\"ur Experimentalphysik, Luruper Chaussee 149, D 22761 Hamburg, Germany\\
\inst{3}Max-Planck-Institut f\"ur Kernphysik, P.O. Box 103980, D 69029 Heidelberg, Germany\\
\inst{4}Dublin Institute for Advanced Studies, 31 Fitzwilliam Place, Dublin 2, Ireland\\
\inst{5}National Academy of Sciences of the Republic of Armenia,  Marshall Baghramian Avenue, 24, 0019 Yerevan, Republic of Armenia\\
\inst{6}Yerevan Physics Institute, 2 Alikhanian Brothers St., 375036 Yerevan, Armenia\\
\inst{7}Institut f\"ur Physik, Humboldt-Universit\"at zu Berlin, Newtonstr. 15, D 12489 Berlin, Germany\\
\inst{8}University of Namibia, Department of Physics, Private Bag 13301, Windhoek, Namibia\\
\inst{9}GRAPPA, Anton Pannekoek Institute for Astronomy, University of Amsterdam,  Science Park 904, 1098 XH Amsterdam, The Netherlands\\
\inst{10}Department of Physics and Electrical Engineering, Linnaeus University,  351 95 V\"axj\"o, Sweden\\
\inst{11}Institut f\"ur Theoretische Physik, Lehrstuhl IV: Weltraum und Astrophysik, Ruhr-Universit\"at Bochum, D 44780 Bochum, Germany\\
\inst{12}GRAPPA, Anton Pannekoek Institute for Astronomy and Institute of High-Energy Physics, University of Amsterdam,  Science Park 904, 1098 XH Amsterdam, The Netherlands\\
\inst{13}Institut f\"ur Astro- und Teilchenphysik, Leopold-Franzens-Universit\"at Innsbruck, A-6020 Innsbruck, Austria\\
\inst{14}School of Physical Sciences, University of Adelaide, Adelaide 5005, Australia\\
\inst{15}LUTH, Observatoire de Paris, PSL Research University, CNRS, Universit\'e Paris Diderot, 5 Place Jules Janssen, 92190 Meudon, France\\
\inst{16}Sorbonne Universit\'es, UPMC Universit\'e Paris 06, Universit\'e Paris Diderot, Sorbonne Paris Cit\'e, CNRS, Laboratoire de Physique Nucl\'eaire et de Hautes Energies (LPNHE), 4 place Jussieu, F-75252, Paris Cedex 5, France\\
\inst{17}Laboratoire Univers et Particules de Montpellier, Universit\'e Montpellier, CNRS/IN2P3,  CC 72, Place Eug\`ene Bataillon, F-34095 Montpellier Cedex 5, France\\
\inst{18}DSM/Irfu, CEA Saclay, F-91191 Gif-Sur-Yvette Cedex, France\\
\inst{19}Astronomical Observatory, The University of Warsaw, Al. Ujazdowskie 4, 00-478 Warsaw, Poland\\
\inst{20}Aix Marseille Universit\'e, CNRS/IN2P3, CPPM UMR 7346,  13288 Marseille, France\\
\inst{21}Instytut Fizyki J\c{a}drowej PAN, ul. Radzikowskiego 152, 31-342 Krak{\'o}w, Poland\\
\inst{22}Funded by EU FP7 Marie Curie, grant agreement No. PIEF-GA-2012-332350, \\
\inst{23}School of Physics, University of the Witwatersrand, 1 Jan Smuts Avenue, Braamfontein, Johannesburg, 2050 South Africa\\
\inst{24}Laboratoire d'Annecy-le-Vieux de Physique des Particules, Universit\'{e} Savoie Mont-Blanc, CNRS/IN2P3, F-74941 Annecy-le-Vieux, France\\
\inst{25}Landessternwarte, Universit\"at Heidelberg, K\"onigstuhl, D 69117 Heidelberg, Germany\\
\inst{26}Universit\'e Bordeaux, CNRS/IN2P3, Centre d'\'Etudes Nucl\'eaires de Bordeaux Gradignan, 33175 Gradignan, France\\
\inst{27}Oskar Klein Centre, Department of Physics, Stockholm University, Albanova University Center, SE-10691 Stockholm, Sweden\\
\inst{28}Wallenberg Academy Fellow, \\
\inst{29}Institut f\"ur Astronomie und Astrophysik, Universit\"at T\"ubingen, Sand 1, D 72076 T\"ubingen, Germany\\
\inst{30}Laboratoire Leprince-Ringuet, Ecole Polytechnique, CNRS/IN2P3, F-91128 Palaiseau, France\\
\inst{31}APC, AstroParticule et Cosmologie, Universit\'{e} Paris Diderot, CNRS/IN2P3, CEA/Irfu, Observatoire de Paris, Sorbonne Paris Cit\'{e}, 10, rue Alice Domon et L\'{e}onie Duquet, 75205 Paris Cedex 13, France\\
\inst{32}Univ. Grenoble Alpes, IPAG,  F-38000 Grenoble, France \protect\\ CNRS, IPAG, F-38000 Grenoble, France\\
\inst{33}Department of Physics and Astronomy, The University of Leicester, University Road, Leicester, LE1 7RH, United Kingdom\\
\inst{34}Nicolaus Copernicus Astronomical Center, ul. Bartycka 18, 00-716 Warsaw, Poland\\
\inst{35}Institut f\"ur Physik und Astronomie, Universit\"at Potsdam,  Karl-Liebknecht-Strasse 24/25, D 14476 Potsdam, Germany\\
\inst{36}Friedrich-Alexander-Universit\"at Erlangen-N\"urnberg, Erlangen Centre for Astroparticle Physics, Erwin-Rommel-Str. 1, D 91058 Erlangen, Germany\\
\inst{37}DESY, D-15738 Zeuthen, Germany\\
\inst{38}Obserwatorium Astronomiczne, Uniwersytet Jagiello{\'n}ski, ul. Orla 171, 30-244 Krak{\'o}w, Poland\\
\inst{39}Centre for Astronomy, Faculty of Physics, Astronomy and Informatics, Nicolaus Copernicus University,  Grudziadzka 5, 87-100 Torun, Poland\\
\inst{40}Department of Physics, University of the Free State,  PO Box 339, Bloemfontein 9300, South Africa\\
\inst{41}Heisenberg Fellow (DFG), ITA Universit\"at Heidelberg, Germany \\
\inst{42}GRAPPA, Institute of High-Energy Physics, University of Amsterdam,  Science Park 904, 1098 XH Amsterdam, The Netherlands\\
\inst{43}Department of Physics, Rikkyo University, 3-34-1 Nishi-Ikebukuro, Toshima-ku, Tokyo 171-8501, Japan\\
\inst{44}Now at Santa Cruz Institute for Particle Physics and Department of Physics, University of California at Santa Cruz, Santa Cruz, CA 95064, USA
}

\begin{document}

\date{Accepted .... Received ...; in original form ...}

\pagerange{\pageref{firstpage}--\pageref{lastpage}} \pubyear{2016}

\label{firstpage}

\maketitle

\footnotetext[1]{Corresponding authors. Correspondence should be sent to \href{mailto:contact.hess@hess-experiment.eu}{contact.hess@hess-experiment.eu}}
\footnotetext[2]{Deceased}

\begin{abstract}
\pks\ ($z=0.055$) was observed with the four H.E.S.S. telescopes  in 2012 during 5.5\,hours. The source was detected above an energy threshold of 200 GeV at a  significance level of $6.1\sigma$. No significant variability is found in these observations. 
The source is well described with a power-law spectrum with photon index  $\Gamma=2.84\pm0.50_{stat}\pm0.10_{syst}$ and normalization (at $E_0$=1.0\,TeV)  $N_0(E_0)=(0.58\pm0.22_{stat}\pm0.12_{syst})\times10^{-12}$\,TeV$^{-1}$cm$^{-2}$s$^{-1}$.
Multi-wavelength data collected with \textit{Fermi}-LAT, \textit{Swift}-XRT, \textit{Swift}-UVOT, ATOM and WISE are also analysed. 
Significant variability is observed only in the \textit{Fermi}-LAT $\gamma$-ray and \textit{Swift}-XRT X-ray energy bands.
Having a good multi-wavelength coverage from radio to very high energy, we performed a broadband modelling from two types of emission scenarios.
The results from a one zone lepto-hadronic, and a  multi-zone leptonic models are compared and discussed. On the grounds of energetics, our analysis favours a leptonic multi-zone model. Models associated to the X-ray variability constraint supports previous results suggesting a BL Lac nature of \pks\, with, however, a large-scale jet structure typical of a radio galaxy.

\end{abstract}

\clearpage
\section{Introduction}
\pks\ (OH\,342) (RA = 06$^\mathrm{h}$27$^\mathrm{m}$06.7$^\mathrm{s}$ DEC = -35$^\circ$29'15'', J2000) is a source located in the  Cluster Abell 3392, observed at a redshift of $z = 0.055$  \citep{Jones2009}. 
The galaxy hosts a supermassive black-hole with mass of $10^{9.19 \pm 0.37}$ M$_{sun}$, deduced by \cite{Bettoni_2003} from bulge luminosity and stellar velocity dispersion relations.
The nature of \pks\ is still a matter of debate, since it reveals  features which appear both radio galaxy and blazar-like in nature.

\cite{Ekers1989} reported the discovery of one-sided jet with a position angle of $+$160$^{\circ}$ and radio halo observed at 5\,GHz with VLA (Very Large Array) extending up to  4\,arcmin.
Radio observations with  VLBA (Very Long Baseline Array) performed at 5\,GHz revealed a strong central core and  a quite faint radio component located at the southern-east side  of the core, consistent with the direction of the large-scale jet orientation \citep{Fomalont2000}.
A bright core emission and a one-sided jet, with orientation of $+$150$^{\circ}$, have been found by \cite{Venturi2000} in VLBI (Very Long Baseline Interferometry)  observations  at 2.3\,GHz.
More recent radio monitoring performed as  part of  the TANAMI (Tracking Active Galactic Nuclei with Austral Milliarcsecond Interferometry)  program has shown  a radio-jet structure  on intermediate scales out to ~95 mas from the core,  corresponding to a distance of 143\,pc \citep{Ojha2010}. 

Using radio observations of \pks\ an estimation of the maximal possible viewing angle of $\theta \leq 61$ deg is obtained from jet to counter-jet ratio, and $\theta \leq 43$ deg from the square root of the ratio of jet power to the jet power for a source at $60$ deg, as defined in \cite{Giovannini_1994}.

Spectroscopic optical observations of \pks\ have revealed that the source  could be a BL Lac object due to the results of the 4000\,\AA~break and  the fit to the optical continuum \citep{Wills2004}.

X-ray BeppoSAX observations in the energy range of 0.2$-$10\,keV of \pks\ have shown a non-thermal hard X-ray component ($\Gamma$=1.7) with a luminosity of $1.8 \times 10^{43}$\,erg\,s$^{-1}$, likely originated in the central object  \citep{Trussoni99}. 
More recent  studies focusing on \textit{Suzaku} and \lat\ observations have revealed to posses a soft spectrum in the X-ray regime and hard in the $\gamma$-ray range 
\citep{Fukazawa_2015}. 
The authors present  broadband spectral modelling with a low Lorentz factor, typical to that usually found for FR\,I radio-galaxies.

In the high-energy band the source has been detected with the \lat\  from 11 months of observations,  and was announced in the Fermi Large Area Telescope First Source Catalogue (1FGL) \citep{1FGL}.
The published \lat\ third source catalogue  reports a flux from \pks\ of $F_{1\mathrm{GeV}-100\mathrm{GeV}} = (1.43 \pm 0.11) \times 10^{-9}$\,ph\,cm$^{-2}$\,s$^{-1}$ and  the hard spectrum characterized by the power-law distribution with the photon index 
$\Gamma_{\mathrm{3FGL}} = 1.88 \pm 0.06$ \citep{3FGL}.

This paper reports the discovery of very high energy $\gamma$-ray emission from the region of \pks. The layout is as follows: Sect.~\ref{hess} introduces  the H.E.S.S. experiment and presents very high energy observations of \pks, Sect.~\ref{data} reports multi-frequency observations of the source, and Sect.~\ref{Sec::Modelling} compares leptonic and lepto-hadronic multi-wavelength (MWL) models, and discusses the physical properties and the source classification.
Hereafter, we adopt a cosmology with $\mathrm{H_0}$ = 71 km s$^{-1}$ Mpc$^{-1}$ , $\Omega_{\Lambda} = 0.73$ and $\Omega_M = 0.27$.

\section{H.E.S.S. observations of \pks} \label{hess}
The High Energy Stereoscopic System (H.E.S.S.) is an array of five Imaging Atmospheric Cherenkov Telescopes (IACTs), located in the Khomas Highland in Namibia,  dedicated to observations of  very-high-energy  $\gamma$ rays (VHE, $E>100$\,GeV), described in detail in \cite{Aharonian2006_crab}. Until 2012, H.E.S.S.I observations were carried out with four 12\,m telescopes, each with a mirror area of  108\,m$^2$, while in 2012 the array was upgraded with a fifth telescope  with a mirror area of  614\,m$^2$. 

\pks\ was observed with H.E.S.S.I four  telescope array during 8 nights in November and December  2012, resulting in a total exposure of 5.5\,h of good quality data. 
All data were taken in wobble mode with the offset of 0.5\,$^{\circ}$. For all the data taken, the observations were between zenith angles of 11-19$^{\circ}$. 

\begin{figure}
\centering{\includegraphics[width=0.49\textwidth]{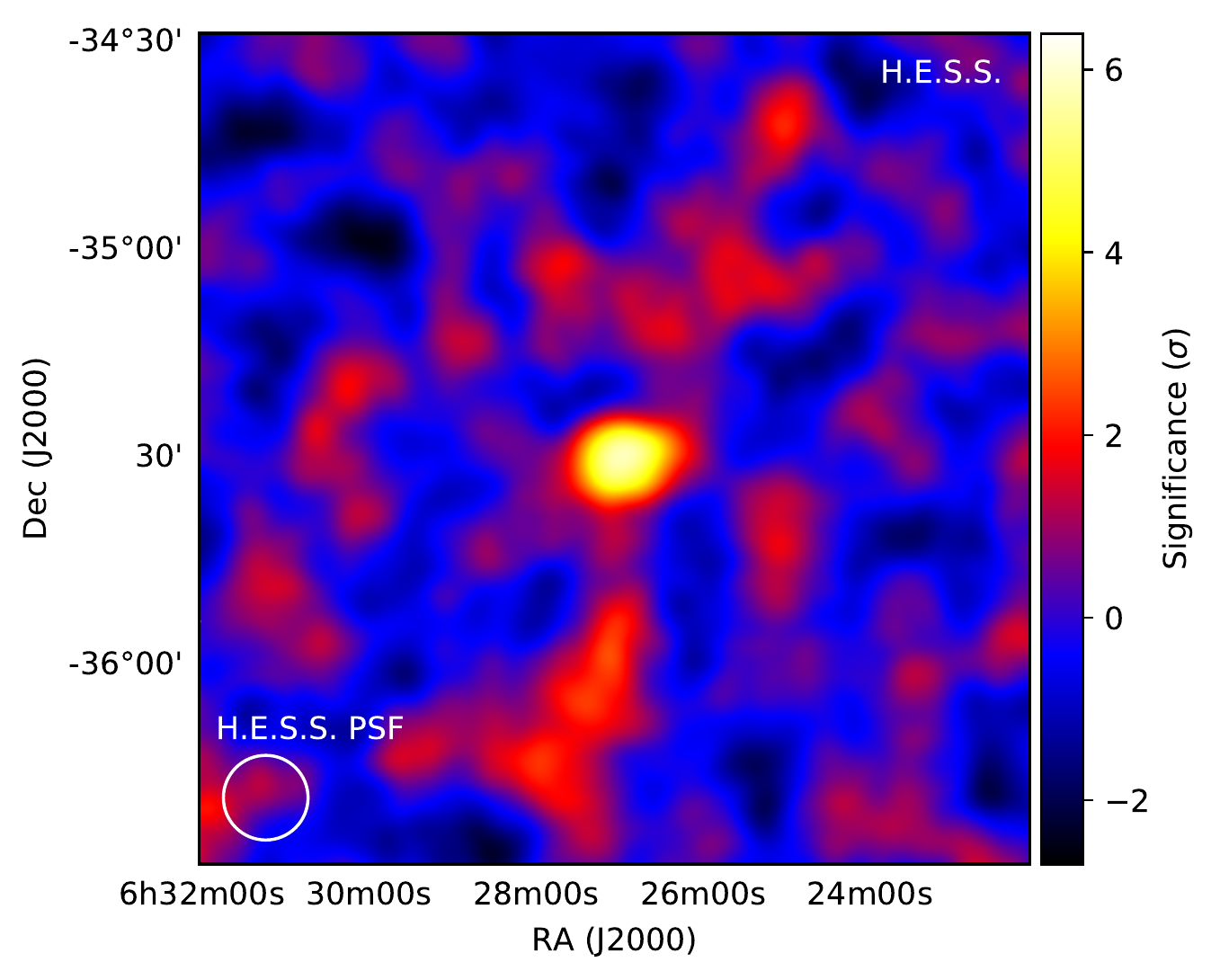}}
\caption[]{H.E.S.S. significance map centred on the position of \pks. The white circle in the bottom left corner shows  the H.E.S.S. point spread function (PSF) as the 68$\%$ containment radius.}
 \label{signif}
\end{figure}

\begin{figure}
\centering{\includegraphics[width=0.49\textwidth]{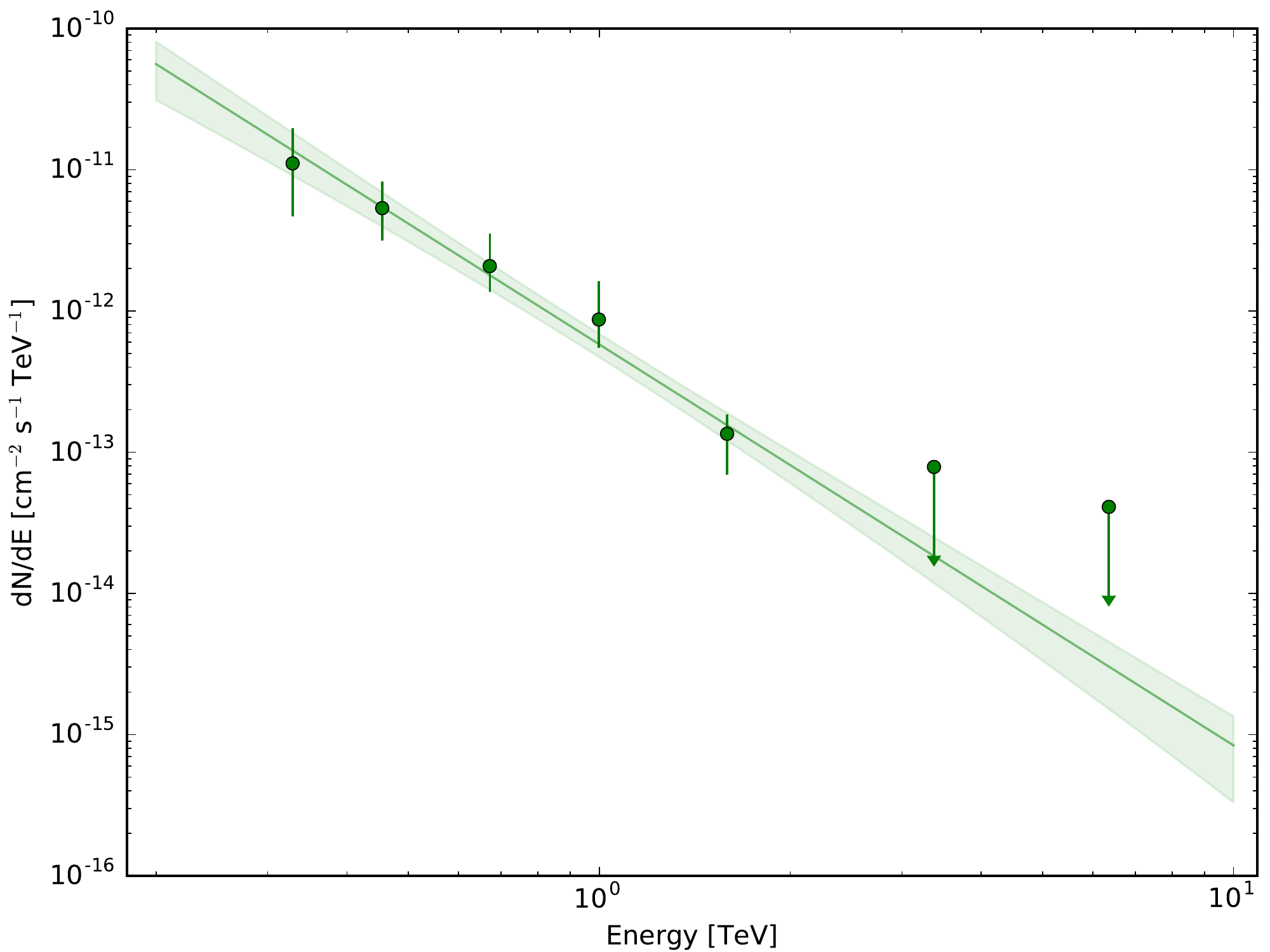}}
\caption[]{H.E.S.S. spectrum of \pks\ above 200\,GeV. The panel presents best power-law fit to the data as a function of the true energy. Upper limits are given at 99$\%$ confidential level with \cite{Feldman98} confidence intervals. The green bow tie area  gives the uncertainty of the fit at the $1\sigma$ confidence
level. }
 \label{hess_spectrum}
\end{figure}

The data were analysed using the Model Analysis chain \citep{Naurois2009} with the \texttt{Loose Cuts} configuration \citep{Crab_paper}. 
The measured excess of 60.7 events corresponds to 6.1$\sigma$ significance \citep[following][]{Li1983}. The emission observed is centred on 
RA = 06$^\mathrm{h}$26$^\mathrm{m}$58.2$^\mathrm{s}$ $\pm$2.6$^\mathrm{s}_{stat}$ DEC = -35$^\circ$29'50''$\pm$50''$\pm$33''$_{stat}$, J2000.
All the results presented in this paper were cross-checked with independently generated calibration files, and an independent analysis
chain called ImPACT \citep{Parsons2014}. 
Figure~\ref{signif} shows the significance map for \pks. Within the statistical uncertainties the source exhibits a
point-like morphology.
The differential energy spectrum of the VHE $\gamma$-ray emission has been derived using a forward folding method \citep{Piron2001} (see Fig.\,\ref{hess_spectrum}). The data are fitted with a power-law model defined as $dN/dE = N_0 (E/E_0)^{-\Gamma}$  with a normalization $N_0(1 TeV)=(0.58\pm0.22_{stat}\pm0.12_{syst})\times10^{-12}$\,TeV$^{-1}$cm$^{-2}$s$^{-1}$, with a photon index $\Gamma=2.84\pm0.50_{stat}\pm0.20_{syst}$. The goodness of the fit is $\chi^2/n_{dof}=24.4/21$. 

The observed integrated flux of \pks\ is $I(>575GeV)=(8.8\pm3.2_{stat}\pm1.6_{syst})\times10^{-13}$\,cm$^{-2}$s$^{-1}$.

The VHE light curve of the nightly integrated flux ($E>200$\,GeV) is presented in the upper panel of Fig.~\ref{lc}. A fit with a constant to the data set yields $\chi^2/n_{dof}=5.95/7$ (p=0.55), indicating no evidence for variability in  VHE $\gamma$-ray emission during the observation period.

\section{Multi-wavelength data} \label{data}

\begin{table*}

\caption[]{Results of spectral fits to \lat\ data using single power-law and log-parabola models. The columns present:
(1) the data model; (2) the normalization (given in 10$^{-13}$\,cm$^{-2}$\,s$^{-1}$\,MeV$^{-1}$); (3) the photon index for the power-law or log-parabola; (4) the curvature parameter; (5) the scale energy in MeV.
}

\centering

\begin{tabular}{c|c|c|c|c}
\hline
\hline
 Model & Normalization & $\Gamma$/$\alpha$ &  $\beta$ &  $E_{0,p}$/$E_{0,l}$ \\
 (1) & (2) & (3) &  (4) & (5) \\ 
 \hline
power-law                  & 3.54$\pm$0.22$_{stat}\pm$0.30$_{syst}$ & 1.87$\pm$0.04$_{stat}\pm$0.01$_{syst}$  & -- &  2000 \\
log-parabola               & 3.74$\pm$0.27$_{stat}\pm$0.30$_{syst}$ & 1.83$\pm$0.05$_{stat}\pm$0.01$_{syst}$  & 0.034$\pm$0.025$_{stat}\pm$0.002$_{syst}$ &  2000 \\
\hline
\end{tabular}
\label{table_fermi_par}

\end{table*}

\subsection{\textit{Fermi}-LAT observations} 
High energy (HE, E$>$100\,GeV) $\gamma$-ray emission from the direction of \pks\ was first reported in the first Fermi-LAT source catalog (1FGL). The source has been also included in the second and third \lat\ source catalog \citep[][2FGL and 3FGL, respectively]{2FGL, 3FGL}.

In this paper, the \lat\ data collected between August 4, 2008 and May 30, 2015  have been analysed using standard \textit{\textup{Fermi Science Tools}} (version v10r0p5) with \verb|P8R2_SOURCE_V6| instrument response functions (IRFs), which is the  latest LAT data release \citep[i.e. Pass\,8; ][]{Atwood2013_pass8}. 
For these studies all photons in the energy range from 100\,MeV to 300\,GeV are selected. 
The maximum zenith angle of $90^\circ$ has been applied. 
The region of interest (ROI) is defined to have  $10^\circ$ size and it is centred on the source. 

The binned maximum-likelihood method \citep{Mattox96} was applied, with 
the Galactic diffuse background  modelled using the \verb|gll_iem_v06| map
cube, and the extragalactic diffuse and residual instrument backgrounds 
modelled jointly using the \verb|iso_P8R2_SOURCE_V6_v06| template. 

All the sources from the 3FGL inside the ROI of \pks\ were included in the model. The residual maps showed an excess in four regions with a significance higher than 5$\sigma$. The model was refined by adding four point-like sources which emission was described by a power-law spectrum. After fitting the spectral parameters of the additional sources the residual maps have been improved and no additional regions with significant excess have been found.
The spectral parameters of \pks\ were fifted simultaneously with those of the Galactic and isotropic emission, and those of the closest sources.

In order to find the best description of  the HE spectrum of \pks\ two models were tested:
\begin{itemize}
\item a power-law in the form of $dN/dE = N_p (E/E_{0,p})^{-\Gamma}$, where $N_p$ is the normalization, $\Gamma$ the photon index, and $E_{0,p}$ the scale energy parameter, 
\item a logarithmic parabola in the form of $dN/dE = N_l (E/E_{0,l})^{-(\alpha + \beta \log (E/E_{0,l})) }$, where $N_{l}$ is the normalization, $\alpha$ the spectral index, $\beta$ the curvature parameter,  and $E_{0,l}$ the scale energy parameter.
\end{itemize}
  
The fit parameters are collected in Table\,\ref{table_fermi_par}. 
The Test Statistic (TS) value between the log-parabola and power-law model is 2.2, meaning that the fit was not significantly improved. 
For further studies we limit the analysis to the simpler, power-law model.

\begin{figure*}
\centering{\includegraphics[width=0.9\textwidth]{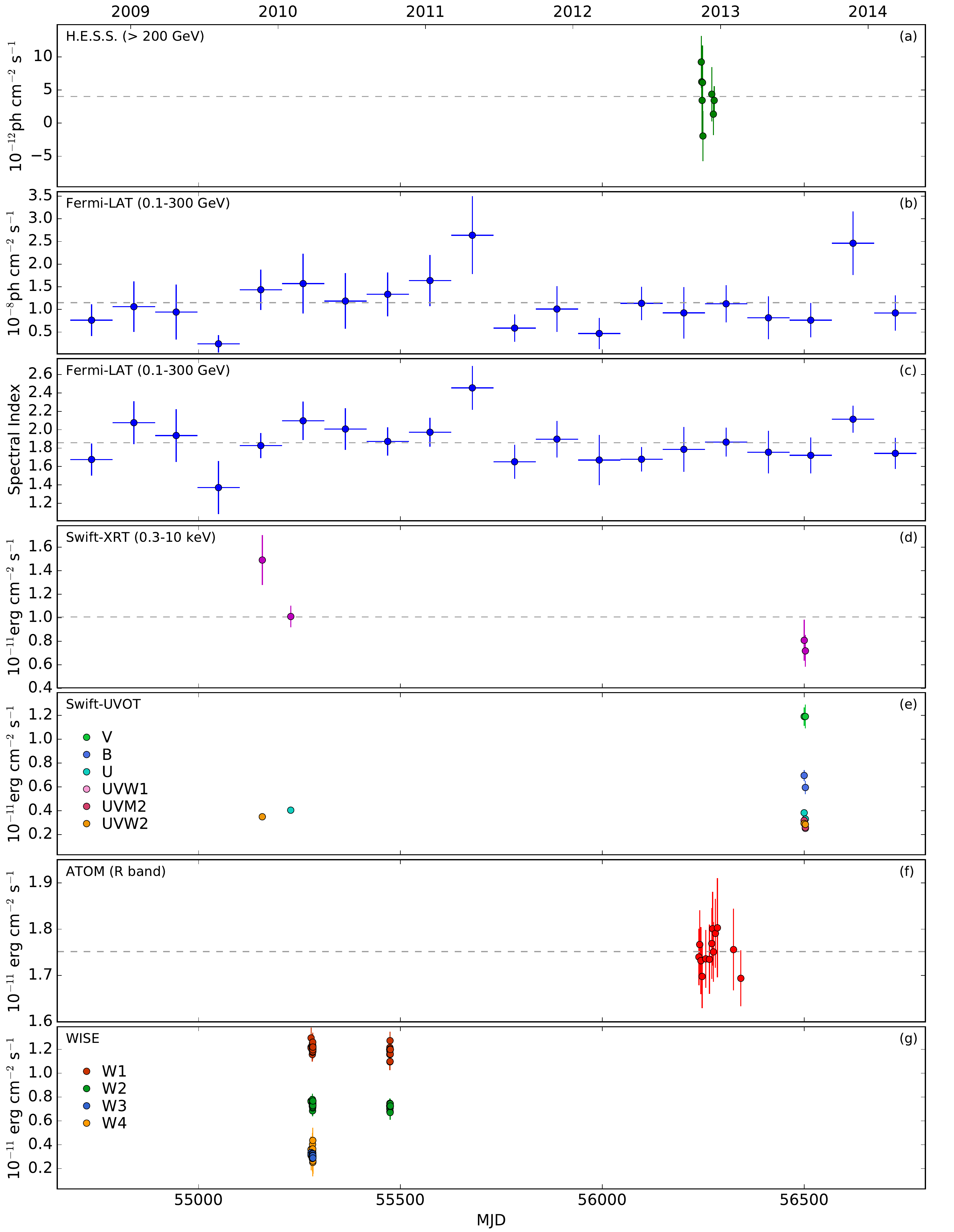}}
\caption[]{Multi-wavelength light curve of \pks. Panels from top to bottom present: H.E.S.S. flux points above 200\,GeV given in night-wise bins; the \lat\ flux  and photon index points in the energy range of 0.1-300\,GeV, each point represents about 120\, days of monitoring; Swift-XRT flux points for the energy range of 0.3-10\,keV binned in one-day intervals; Swift-UVOT flux points  for  V, B, U, UVW1, UVM2, and UVW2 filters, each point represents one day of observations; ATOM optical flux points given in R band; WISE observations in W1--W4 filters. Swift-UVOT, ATOM and WISE observations are corrected for the influence of the extinction.
The horizontal dashed lines in most of the panels represent the average flux for all observations presented in a given band.}
\label{lc}
\end{figure*}

In order to calculate spectral points, the data have been divided into six logarithmically equally spaced energy bins and for each bin a separate likelihood analysis has been run. 
The normalisation of the 3 nearest point sources (one from the 3FGL and two additionnal sources of the four mentioned above) have been let free during the fit, as the spectral parameters of the Galactic and isotropic diffuse emissions. We fix the spectral index of \pks\ to 2 to avoid any dependence on the spectral model found for the whole energy range.
A threshold of $\mathrm{TS} < 9$ (corresponding to a significance of 3 sigma) is imposed in each energy bin for a flux calculation, otherwise an upper-limit is calculated.
A one-$\sigma$ confidence contour has been calculated using the covariance matrix obtained with the \verb|gtlike| procedure (see Fig.\ref{fig::SED_data_PKS0625}).
For the light curves, the same spectral parameters have been let free as for the SED computation, with free spectral index for \pks.
The long term  HE flux and spectral index light curves for \pks\ are presented in the  second and third panels  (b and c) Fig.~\ref{lc}. Each bin corresponds to about 120 days of monitoring.
A fit to a constant for the flux points, not considering systematic errors, yields$\chi^2/n_{dof}=41.4/19$ ($p<0.01$), showing indication for variability in the HE $\gamma$-ray regime. This is also confirmed in the fractional variability amplitude value, calculated following \cite{Vaughan2003},  equal F$_{var}$=75\%.
Also significant variability has been found in the case of  the spectral index time evolution.  A fit with a constant results in  $\chi^2/n_{dof}=34.0/19$ ($p=0.02$).

\begin{table*}

\caption[]{Parameters of the spectral analysis of the \textit{Swift}-XRT data. The  columns present: (1) the observation ID number; (2) the observation date; (3) the time exposure given in seconds; (4) the flux observed in the energy range of 0.3-10\,keV in  units of 10$^{-11}$\,erg\,cm$^{-2}$\,s$^{-1}$; (5) the spectral index for the power-law fit to data; (6) $\chi^2$ statistics value and the number of degrees of freedom. }
\centering
\begin{tabular}{c|c|c|c|c|c}
\hline
\hline
Observation ID & Observation date & Exposure  &  F$_{0.3-10\,keV}$   &   $\Gamma$ & $\chi^2/dof$    \\
(1) & (2) & (3) &  (4) &   (5) & (6)    \\ 
\hline
 00039136001      & 22/11/2009 & 1384.1  & 1.49$^{+0.21}_{-0.19}$  & 2.079$\pm$0.093 & 8.64/10   \\
 00039136002      & 01/02/2010 & 4550.2  & 1.01$^{+0.10}_{-0.11}$ & 2.271$\pm$0.051 & 38.24/30  \\
 00049667001      & 26/07/2013 & 8110.6  & 0.81$^{+0.19}_{-0.17}$  & 2.125$\pm$0.041 & 49.68/43  \\
 00049667002      & 29/07/2013 & 2462.1  & 0.72$^{+0.14}_{-0.14}$  & 2.069$\pm$0.084 & 12.54/10  \\
 All observations &            & 16506.0 & 0.81$^{+0.20}_{-0.19}$  & 2.149$\pm$0.027 & 88.82/87  \\
 \hline
\end{tabular}
\label{table_xrt}

\end{table*}

\begin{table*}

\caption[]{Magnitudes for different epochs from  the \textit{Swift}-UVOT data. The  columns  present: (1) the observation ID number  and (2)--(7) the observed magnitudes in V, B, U, UVW1, UVM2 and UVW2 bands, respectively. The magnitudes  
are not corrected for the Galactic extinction.  The hyphen (--) indicates that the were no observation taken in a given filter for a specific observation ID.}
\centering

\begin{tabular}{c|c|c|c|c|c |c}
\hline
\hline
 Observation ID & V & B &  U &  UVW1 & UVW2 & UVM2    \\
 (1) & (2) & (3) &  (4) &   (5) & (6)  & (7)  \\ 
 \hline
 00039136001       & --             & --             & --             & --             & 16.29$\pm$0.06  & -- \\
 00039136002       & --             & --             & 16.21$\pm$0.05 & --             & --              & -- \\
 00049667001       & 15.57$\pm$0.07 & 16.52$\pm$0.07 & 16.27$\pm$0.08 & 16.37$\pm$0.06 & 16.48$\pm$0.07  & 16.29$\pm$0.06 \\
 00049667002       & 15.57$\pm$0.09 & 16.69$\pm$0.10 & 16.43$\pm$0.12 & 16.54$\pm$0.12 & 16.51$\pm$0.08  & 16.51$\pm$0.12 \\
 Average magnitude & 15.57$\pm$0.07 & 16.52$\pm$0.07 & 16.34$\pm$0.05 & 16.38$\pm$0.06 & 16.47$\pm$0.07  & 16.29$\pm$0.06 \\
\hline
\end{tabular}
\label{table_uvot}

\end{table*}

\subsection{\textit{Swift}-XRT and UVOT observations}
X-ray observations of \pks\ were performed with  the XRT detector on-board the \textit{Swift} spacecraft \citep{Burrows20015}. The source was observed in  photon counting (PC) mode in the energy range of 0.3-10\,keV in four pointing observations with  a total exposure of 16.5\,ks. 
All the observations were analysed with the HEASOFT version\,6.15 package\footnote{\url{http://heasarc.gsfc.nasa.gov/docs/software/lheasoft}} following the standard \verb|xrtpipeline| procedure. 
The spectral analysis was performed with  the \verb|XSPEC| package (version 12.8.2). A circular region with a radius of 5''  around the position of the source was used. The same size off region was used in order to determine the background. 
The logarithmic energy bin sizes were adopted so as to ensure a minimum count of 20 events per bin. The spectra were fitted with a single power-law function with a Galactic  absorption 
 value of $N_{H} = 6.5 \times 10^{20}$\,cm$^{-2}$ taken from \cite{Kalberla05} which was set as a frozen parameter.  All  measured \textit{Swift}-XRT fluxes are collected in Table\,\ref{table_xrt}.

Simultaneously with XRT, \pks\ was observed with the UVOT instrument \citep{Roming2005} on-board the \textit{Swift} spacecraft in six optical and ultraviolet (UV) filters, namely: U (345\,nm), B (439\,nm), V (544\,nm), UVW1 (251\,nm), UVM2 (217\,nm) and UVW2 (188\,nm).  The magnitudes and corresponding fluxes were calculated using the \verb|uvotmaghist| tool and the conversion factors provided by \cite{Poole08}. 
The optical and ultraviolet observations were corrected for  dust absorption using the reddening coefficient $E(B-V) = 0.0562$\,mag \citep{Schlafly} and  the ratios of the extinction to reddening, $A_{\lambda} / E(B-V)$, for each filter from \cite{Giommi2006}.  The magnitudes  are collected in Table\,\ref{table_uvot}.

The optical XRT and UVOT light curves are presented in the Fig.~\ref{lc}\,d-e. 
In the X-ray regime, the first observation represents a significantly higher state of the source than in the case of the three latter data points. 
 The limited number of UVOT observations does not allow either  claiming or  excluding  source variability in the optical and ultraviolet bands. However in the optical and ultraviolet UV bands there is no indication for an elevated flux level corresponding with the one found in the X-ray observations in November 2009.
  The X-ray bright state is not associated with a simultaneous brightening at lower energies, even though the X-ray spectral index of about 2 (consistent
with the average value) would suggest an achromatic variability of the
synchrotron component. This may indicate that the UV and X-ray photons do not share a common origin.

\subsection{ATOM observations}
ATOM (Automatic Telescope for Optical Monitoring) is a 75\,cm optical telescope located in Namibia near the H.E.S.S. site \citep{Hauser2004}.
\pks\ was observed with ATOM only in the  R band in 13  observations. 
The data collected have been analysed using an aperture of 4'' radius and differential photometry. The observations have been corrected for dust absorption using the absorption magnitude $A_R = 0.144$\,mag from  \citet{Schlafly}.
The optical ATOM light curve is presented in Fig.~\ref{lc}f. The long term ATOM observations do not indicate variability during the period presented in the paper. A fit with a constant to the data results in $\chi^2/n_{dof}=2.75/12$ ($p=0.997$).

\subsection{WISE observations}
Wide-field Infrared Survey Explorer (WISE) is a space telescope which performs observations in the infrared energy band at four wavelengths: 3.4\,$\mu$m (W1), 4.6\,$\mu$m (W2), 12\,$\mu$m (W3) and 22\,$\mu$m (W4). The spectral data are taken from the AllWISE Source Catalog and the light curve from the AllWISE Multi-epoch Photometry Table\footnote{\url{http://wise2.ipac.caltech.edu/docs/release/allwise/}}. The magnitudes are converted to flux by applying the standard procedure \citep{Wright_2010} and for W1 and W2 we apply the colour correction to a power-law with a photon index of 0, as suggested for  Galactic emission. Since W3 and W4 are widely dominated by the non-thermal radiation and show similar flux densities, we use the colour correction to a power-law with a  photon index of -2.
The infrared light curves do not show  variability within  uncertainties of the measurements. For each of the two observational sessions the reduced $\chi^2$ of a fit with a constant value is lower than 0.5 for a given filter.

\section{Modelling}
\label{Sec::Modelling}

\subsection{Multi-wavelength spectral energy distribution}

\begin{figure*}
\centering{\includegraphics[width=0.9\textwidth]{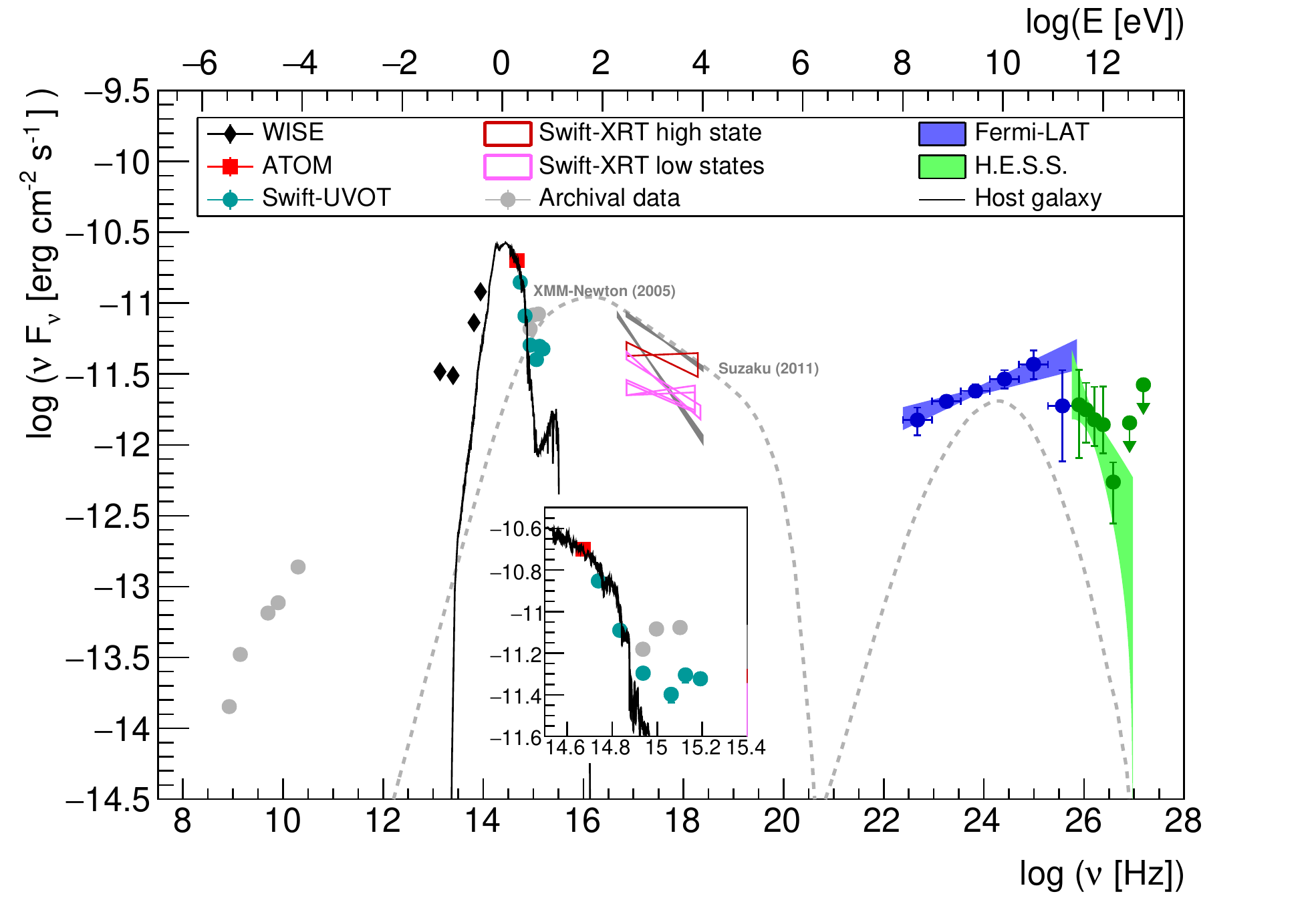}}
\caption{Multi-wavelength SED.
 The black line is the simulated emission of the host galaxy seen in the UVOT aperture of 5". The dotted grey lines are synchrotron and inverse-Compton emission from a leptonic synchrotron-self-Compton (SSC) model, with a variability timescale of $10^5$ s performed in \protect\cite{Fukazawa_2015}. An optical-UV zoom of the SED is presented in the inset.}
\label{fig::SED_data_PKS0625}
\end{figure*}

The data from H.E.S.S., \textit{Fermi}-LAT, \textit{Swift}-XRT, \textit{Swift}-UVOT, ATOM, and WISE have been used together to build the global multi-wavelength spectral energy distribution (SED) of the source in addition to archival radio data from the Australia Telescope 20-GHz Survey \citep[AT20G,][]{Murphy_2010}, NRAO VLA Sky Survey \citep[NVSS,][]{Condon_1998} and Sydney University Molonglo Sky Survey \citep[SUMSS,][]{Mauch_2003} catalogues. 
In order to have a deeper look at the source variability, UV and X-ray data from \textit{XMM}-Newton in 2005 \citep{Gliozzi_2008}, and an X-ray spectrum from Suzaku in 2011 \citep{Fukazawa_2015}, are also represented in the SED (see Figure \ref{fig::SED_data_PKS0625}).
Two states of the multiple \textit{Swift}-XRT observations are differentiated, one high state observed on November 22, 2009, and low states from February 01, 2010 to July 29, 2013.
The lack of frequent X-ray monitoring leaves room for the possibility of a fast X-ray flare on November 22, 2009, which in such a case could have been missed by \textit{Fermi}-LAT and thus would be inconsistent with the high energy spectra used for the modelling. In the absence of evidence for such a fast event, we assume for the following MWL models that the measured X-ray variability does not significantly impact the \textit{Fermi}-LAT and H.E.S.S. spectra.

The prominent IR-optical bump of the SED is due to the host galaxy emission. In order to avoid a misidentification of the non-thermal flux density at these energies a separate host galaxy contribution is  included in the model fit.
Following \cite{Inskip_2010}, the host galaxy presents an effective radius of $18.77 \pm 0.98$" for a flux density of $6.00\times 10^{-11}$~erg~cm$^{-2}$~s$^{-1}$ in the $\mathrm{K_s}$ band.
By simulating the emission from a giant elliptical galaxy, as suggested in \cite{Wills2004}, with PEGASE 2 templates \citep{Fioc_1999}, we deduce an associated galaxy mass of $M_{\mathrm{host}} = 9.15\times 10^{11} M_{\odot}$.

Having the black hole mass $\log({M_{\bullet} / M_{\odot}}) = 9.19 \pm 0.37$ \citep{Bettoni_2003}, we can check the consistency of the deduced host stellar mass with the one expected by the empirical relation of broad-line AGN from \cite{Reines_2015}, expressed as
\begin{equation}
\label{eq::BH_mass}
\log({M_{\bullet} / M_{\odot}}) = \alpha + \beta \log({M_{\mathrm{host}} / 10^{11} M_{\odot}}),
\end{equation}
with $\alpha = 7.45 \pm 0.08$ and $\beta = 1.05 \pm 0.11$.

The associated host mass is then $\log({M_{\mathrm{host}} / M_{\odot}}) = 12.66^{+0.67}_{- 0.55}$. The lower limit of this result is slightly bigger than our value with a factor 1.4. We can however note that the relation \ref{eq::BH_mass} is derived from sources with significantly lower black holes masses than the one of \pks ~($\log({M_{\bullet} / M_{\odot}}) < 8.2$), leading to a possible bias of this estimation.

We  report the galaxy emission in the SED as seen in the UVOT filters with an aperture of 5", this emission is hence underestimated in WISE filters (W1, W2) having  much  wider apertures.
As seen in Figure \ref{fig::SED_data_PKS0625}, this model of the host galaxy is in good agreement with the UVOT and ATOM data, where the optical points are dominated by the host.

The difference in luminosity between \textit{Swift}-UVOT and \textit{XMM}-Newton observations is indicative of  significant variability in the UV range. The UV data from \textit{XMM}-Newton are also suggestive of a hard spectrum. Considering the strong excess due to the host galaxy at low frequencies, the intrinsic spectrum should be much harder than that  observed. This indicates a sharp spectral variation in the UV band.
The X-ray data indicate strong variability in both flux and photon index. High and low states both reveal various photon indexes throughout the observation period, making difficult the deduction of a variability pattern.

One-zone modelling of the source emission was recently discussed by \cite{Fukazawa_2015}, presented in Figure \ref{fig::SED_data_PKS0625}. They found parameters compatible with the source being  radio-galaxy in nature, such as a low Doppler factor.
We note however that the only weak observational constraints were available for such conclusions. 
Indeed, their model takes into account \textit{XMM}-Newton UV data from 2005 and Suzaku X-ray data from 2011. Considering the complex UV and X-ray variabilities, these data from different periods cannot be safetly used for a general broadband emission model.
With the new dataset presented in this paper we are able to provide stronger constraints for MWL emission scenarios.

\subsection{Leptonic scenario}
\label{Sec::Leptonic}

With the WISE data showing no variability for  observing periods, we made the strong assumption of a constant IR flux for the source. As seen in Figure \ref{fig::SED_data_PKS0625}, luminosities measured in WISE W4 and W3 bands cannot be associated to  thermal emission from the host galaxy or a dusty torus, due to their low frequencies. These bands are therefore considered to be dominated by synchrotron emission.
No hints of multi-wavelength variation associated with the two \textit{Swift}-XRT X-ray states can be deduced from the observations (see Figure \ref{lc}). The steady state of the UVM2 band simultaneously with two X-ray states can lead to at least two possible interpretations: 
\begin{itemize}
\item[$\bullet$] Two non-correlated non-thermal components are radiating from far-IR (WISE W4 and W3 bands) to X-rays bands.
\item[$\bullet$] There is only one component from far-IR to X-rays and the UVM2 band is precisely at the tipping point of a spectral change of the synchrotron emission between the two states.
\end{itemize}
The second interpretation is not appropriate for a one-zone leptonic synchrotron-self-Compton (SSC) scenario. It would induce a very wide gap between a synchrotron peak in the IR-optical range and the inverse-Compton peak at the energy of about $100$ GeV, leading to unconventional parameters such as an unusually low magnetic field and a wide emitting region resulting in  emission being far from equipartition between magnetic field and non-thermal particles. Moreover, the synchrotron bump would be much wider than the inverse-Compton one with similar peak luminosities. Such asymmetry is not describable by a pure one-zone SSC scenario \citep[similar case of SED as AP Librae;][]{hess_ap}.

\begin{figure}
\centering{\includegraphics[width=0.49\textwidth]{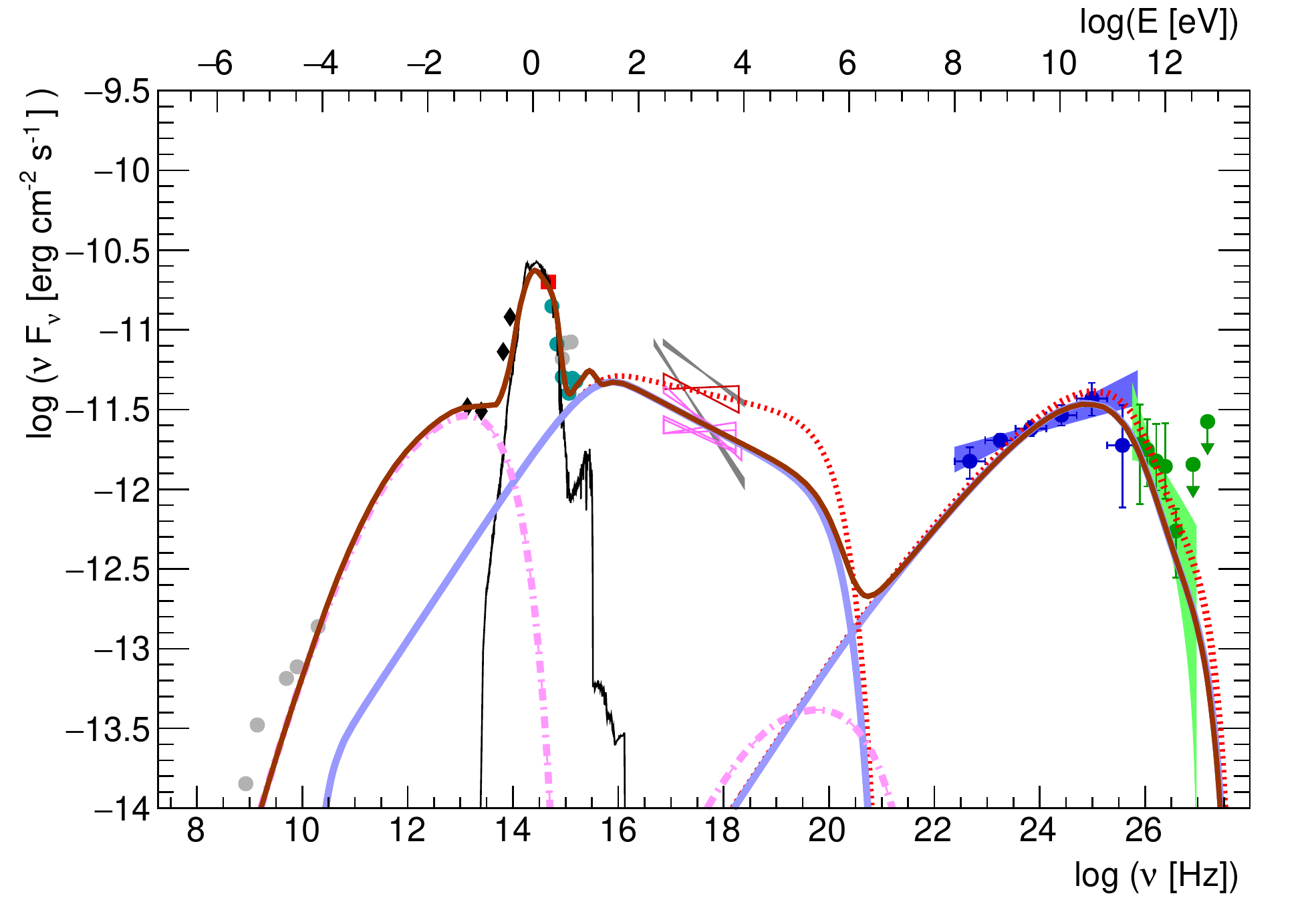}}
\caption{Leptonic multi-zone SSC modelling of the SED. The blob synchrotron and SSC emissions are represented in blue for the low state and dotted red for the high state. The magenta dotted-dashed lines are the jet synchrotron and SSC emissions. The EBL absorption is taken into account following the description of \protect\cite{Finke_2010}.}
\label{fig::SED_PKS0625_multizone}
\end{figure}

Hence, in this first approach we consider the presence of two SSC components, following the 'blob-in-jet' description given in \cite{Katarzynski_2001}, and completed by \cite{Hervet_2015}. The main radiating component is a Doppler boosted spherical compact blob, composed of a turbulent magnetic field and  e$^{+}$e$^{-}$ non-thermal population following a broken power-law energy spectrum. 
The second component is the leptonic conical stratified base of the extended jet surrounding (or in front of) the blob, with its particle energy spectrum defined by a simple power-law. A similar stratified conical jet model was also described by \cite{Potter_2012}.
We consider its synchrotron emission dominating the radio-to-IR luminosity of the source.
The high radio flux and the bright kpc jet structure of \pks\  does supports such a scenario. 
The resulting multi-wavelength model is presented in Figure \ref{fig::SED_PKS0625_multizone}, with the associated parameters adopted being provided in Table \ref{table::Params_multizone}.
The modelling provides a good multi-wavelength representation for both states, with the radio-to-IR emission described by the synchrotron radiation of the jet and a blob synchrotron peak in the UV--X-ray range.
Hence, the two XRT states can be modelled by changing only the second slope index of the electron energy spectrum, from $n_2 = 3.15$ to $n_2 = 3.35$, which has only a minor effect on the spectrum in the high-energy bump. 
However, a natural explanation of index variation remains unclear.

\begin{table}
\caption{Values of parameters used for the leptonic multi-zone SSC model shown in Figure \ref{fig::SED_PKS0625_multizone}. $\theta$ is the angle of the jet direction with the line of sight, $\Gamma$ expresses the Lorentz factor, $K$ is the normalisation factor of the particle density, $n_1$ and $n_2$ are the first and second slope of the blob electron spectrum, $R$ is the radius of the blob, $B_1$ and $R_1$ are the magnetic field and the radius of the first the stratified jet slice respectively. $L$ is the total length of the jet, $\alpha/2$ is the semi-aperture angle of the jet. Parameters marked with asterisk (*) are considered in the host galaxy frame (see \protect\cite{Katarzynski_2001,Hervet_2015} for a detailed description of the model).}
\centering
\begin{tabular}{ccc}
        \hline \hline
        \noalign{\smallskip}
    Parameter & Value & Unit\\ \hline
    \noalign{\smallskip}
    $\theta$ & $1.0$ & deg\\ 
    \hline
    Blob\\
    \hline
    $\Gamma$ & $10.4$ & $-$\\ 
    $K$ & $2.3\times 10^{3}$ & cm$^{-3}$\\ 
    $n_1$ & $2.0$ & $-$\\
    $n_2$ (low state) & $3.35$ & $-$\\
    $n_2$ (high state) & $3.15$ & $-$\\
    $\gamma_{\mathrm{min}}$ & $1.0$ & $-$\\
    $\gamma_{\mathrm{max}}$ & $6.0\times 10^{6}$ & $-$\\
    $\gamma_{b}$ & $4.0\times 10^{4}$ & $-$\\
    $B$ & $4.0\times 10^{-2}$ & G\\
    $R$ & $9.0\times 10^{15}$ & cm\\
    \hline
    Jet\\
    \hline
    $\Gamma$ & $4.1$ & $-$\\ 
    $K$ & $8.5\times 10^{2}$ & cm$^{-3}$\\ 
    $n$ & $2.1$ & $-$\\
    $\gamma_{\mathrm{min}}$ & $1.0$ & $-$\\
    $\gamma_{\mathrm{max}}$ & $3.2\times 10^{3}$ & $-$\\
    $B_1$ & $3.1\times 10^{-1}$ & G\\
    $R_1$ & $1.5\times 10^{16}$ & cm\\
    $L$* & $3.0\times 10^{2}$ & pc\\
    $\alpha/2$* & $1.0$ & deg\\
    \hline
\hline
\end{tabular}
\label{table::Params_multizone}
\end{table}

The parameters used provide a minimal variability of 4.4 h for the blob and 18.3 h for the jet.
The equipartition parameter deduced from the energetics (see Table \ref{table::Lepto_energetics}) indicates a matter-dominated blob with $P_B/P_e = 2.97\times 10^{-3}$ in the low state and $P_B/P_e = 2.95\times 10^{-3}$ in the high state. The jet emission is considered stationary and close to equipartition, with an output value $P_B/P_e = 0.99$. 
In this blob-in-jet structure, the blob is carrying $\simeq 83 \%$ of the total power ($P_{\mathrm{tot}} \simeq 2.1\times 10^{43}$ erg s$^{-1}$).

\begin{table}
\caption{Source energetics deduced from the leptonic model of jet and blob for the low and high activity states. $P_B$, $P_e$ and $P_R$ are respectively magnetic, non-thermal kinetic and radiating powers. All powers are expressed in [erg s$^{-1}$].}
\centering
\begin{tabular}{lccc}
        \hline \hline
        \noalign{\smallskip}
    Powers & Jet & Blob low & Blob high\\ \hline
    \noalign{\smallskip}
    $P_B$ 	& $1.35 \times 10^{42}$ & $5.20 \times 10^{40}$ & $5.20 \times 10^{40}$\\ 
    $P_e$ 	& $1.36 \times 10^{42}$ & $1.75 \times 10^{43}$ & $1.76 \times 10^{43}$\\
    $P_R$ 	& $4.21 \times 10^{41}$ & $2.19 \times 10^{41}$ & $2.81 \times 10^{41}$\\
    \hline
\hline
\end{tabular}
\label{table::Lepto_energetics}
\end{table}

The SED shape naturally favours a rather high Doppler factor associated to a small angle with the line of sight. However, even by having a less convincing fit and higher energetics, the known parameter degeneracy of SSC models allows a significant increase of the angle, associated to low Doppler factor and wide emission zone. A limit of $\theta \lesssim 15$ deg is, however, placed by when the maximum variability time constraint of 41 days between the two \textit{Swift}-XRT states cannot be sustained anymore with a good MWL representation.

\subsection{Lepto-hadronic scenario} 

As discussed in the previous section, a wide one-zone synchrotron bump from IR to X-rays is not well suited for a leptonic scenario. However this interpretation can be explained if we consider a lepto-hadronic scenario.

In this case we interpret the low energy SED bump as synchrotron radiation from primary relativistic electrons and a high-energy bump strongly dominated by synchrotron radiation from relativistic protons  \citep{Mucke_2000,Aharonian_2000,Mastichiadis_2013,Boettcher_2013,Cerruti_2015}. The following results are based on the model developed by \cite{Boettcher_2013}.

\begin{figure}
\centering{\includegraphics[width=0.49\textwidth]{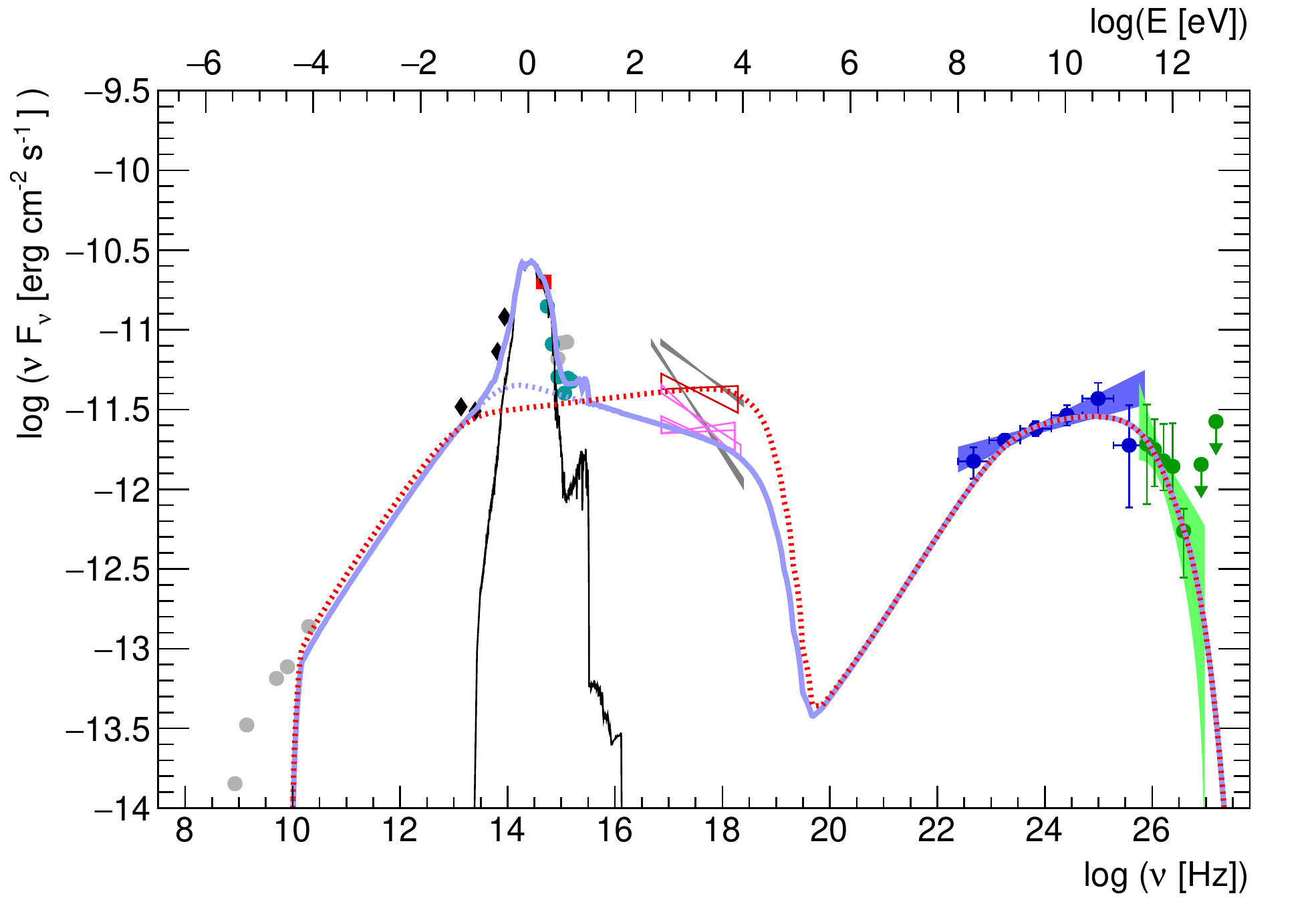}}
\caption{Lepto-hadronic one-zone SSC modelling of the SED. The low state is represented by the blue line and the high state by the red dotted line. The EBL absorption is taken into account following the description by \protect\cite{Finke_2010}.}
\label{fig::SED_PKS0625_hadro_02}
\end{figure}

Contrary to a leptonic SSC scenario, the lepto-hadronic model can naturally produce asymmetric low-energy and high-energy emission bumps. 
Indeed, the energy spectra  may be different, which strongly reduces the constraints given by the SED shape. 
The main parameter differentiating the two states is the index of the electron energy spectrum, which is harder for the high state.  
The ways that a change of the electron spectrum can affect the protons are complex and not included in this model.
Without the observation of X-ray flux variability, 
the parameters of the proton population are assumed as constant.
The resulting multi-wavelength model is presented in Figure \ref{fig::SED_PKS0625_hadro_02}, with the associated parameters adopted being provided in Table \ref{table::Params_one_zone_hadro}.

In order to accelerate protons to these high energies, an extremely large magnetic field of 100 G is needed for this model. This implies that the system is extremely far from equipartition, with $P_B/P_e = (7.9-8.8)\times 10^7$ (resp. high and low states), and $P_B/P_p = 4.29\times 10^4$. This leads to  the requirement of unrealistic large jet power of $P_{\mathrm{jet}} \simeq P_B = 3.8\times 10^{47}$ erg s$^{-1}$, significantly larger than the one required by the leptonic scenario.
We note  that this power is within the common range of what is required by hadronic scenario for multiple sources \citep{Protheroe_2001,Boettcher_2013}

Also, for this model the maximum particle Lorentz factor of protons is about 6 magnitudes greater than that  one of electrons. A more efficient acceleration for protons does make sense as they are subject to much smaller radiative losses. However, despite this,  reaching such a difference between the two populations raise difficulties in interpreting the particle acceleration process.

\begin{table}
\caption{Values of parameters used for the lepto-hadronic one-zone SSC model shown in Figure \ref{fig::SED_PKS0625_hadro_02}. $\Gamma$ is the blob Lorentz factor, {$\theta$} is the angle of the blob direction with the line of sight, {$n_e$} and {$n_p$} are the slope of the electron and proton spectra, {$R$} is the radius of the emitting region, $\eta_{\mathrm{esc}}$ is the escape-time parameter, defining a break in the particle spectra (more details on \protect\cite{Boettcher_2013}).}
\centering
\begin{tabular}{ccc}
        \hline \hline
        \noalign{\smallskip}
    Parameter & Value & Unit\\ \hline
    \noalign{\smallskip}
    General\\
    \hline
    $\Gamma$ & $10.0$ & $-$\\ 
    $\theta$ & $5.74$ & deg\\ 
    $B$ & $1.0\times 10^{2}$ & G\\
    $R$ & $1.0\times 10^{16}$ & cm\\ 
    $\eta_{\mathrm{esc}}$ & 3.0 & $-$\\
    \hline
    Electrons\\
    \hline
    $P_e$ (low state) & $4.28\times 10^{39}$ & erg s$^{-1}$\\
    $P_e$ (high state) & $4.72\times 10^{39}$ & erg s$^{-1}$\\
    $n_e$ (low state)& $2.20$ & $-$\\ 
    $n_e$ (high state)& $1.92$ & $-$\\ 
    $\gamma_{\mathrm{min},e} $(low state) & $1.40\times 10^{2}$ & $-$\\
    $\gamma_{\mathrm{min},e} $(high state) & $6.50\times 10^{2}$ & $-$\\
    $\gamma_{\mathrm{max},e}$ & $4.50\times 10^{4}$ & $-$\\
    \hline
    Protons\\
    \hline
    $P_p$ & $8.75\times 10^{42}$ & erg s$^{-1}$\\
    $n_p$ & $1.90$ & $-$\\ 
    $\gamma_{\mathrm{min},p}$ & $1.07 $ & $-$\\
    $\gamma_{\mathrm{max},p}$ & $2.13\times 10^{10}$ & $-$\\
    \hline
\hline
\end{tabular}
\label{table::Params_one_zone_hadro}
\end{table}

\section{Discussion}
\label{Sec::Discussion}

By working with multiple non-simultaneous data, we modelled the source within the framework of the assumptions we made of its multi-wavelength variability. 
The one-zone lepto-hadronic and multi-zone leptonic models both provide a good multi-wavelength representation to the observed SED.
 Although the SED alone cannot exclude one of these scenarios, the multi-zone leptonic scenario is favoured by the energy budget calculations.
 Significantly less power  by a factor $\sim 2.5 \times10^4$ 
and ratio of  non-thermal electron to B-field power  closer to equipartition is required. 
Moreover, the very steep X-ray spectrum observed in 2005 by \textit{XMM}-Newton favours a synchrotron peak within the UV--X-ray range.

Since the host galaxy is widely dominating the optical spectrum, the slight excess in the UVW2 and UVM2 filter of \textit{Swift}-UVOT (see Figure \ref{fig::SED_data_PKS0625}) is naturally explained in our models by the galactic UV emission. However, the stronger UV excess observed with \textit{XMM}-Newton in 2005 is not expected in the two scenarios presented here. The most probable explanation is a past high luminosity of the nucleus big blue bump emission, also supported by the detection of [OII] and [OIII] lines \citep{Wills2004}.
Due to the lack of observational constraints, a nuclear emission  component (accretion emission reprocessed by the broad line region) is not considered in the models.

The radio-galaxy nature of \pks\ has previously been often discussed \citep{Trussoni99,Wills2004,Nesci_2013,Fukazawa_2015}.
Although the large scale radio structure of the source is typical of an FR~I radio-galaxy \citep{Ekers1989}, there is now an accumulated body of evidence for the BL Lac nature of the nucleus and the pc-scale jet.
We present now a summary of the arguments favouring a blazar, or moderately misaligned blazar nature of the source.
\pks\ was defined as a radio-galaxy emitting in HE $\gamma$ rays by \cite{Abdo_2010b}, based on directional associations between the 15 month Fermi catalogue and the radio catalogues 3CR, 3CRR and  MS4.
The main criterion used here to define a radio-galaxy was the spectral slope with an index $\alpha = (\Gamma-1) \geq -0.5$ ($S_{\nu} \propto \nu^{\alpha}$), which relates to a photon index $\Gamma$, at low frequencies (178-408 MHz).
However the VLBI map from the TANAMI collaboration \citep{Ojha2010} shows a clearly one-sided pc jet with superluminal apparent velocities \citep{Muller_2013} of $\beta_{a} =3 \pm 0.5$~c requiring  a maximum jet misalignment $\theta \leq 37^{+ 7}_{- 5}$ deg.

Having no counterjet detection, its maximum flux is constrained by the lowest radio contour defined at $3 \times$ RMS noise, $F_{cjet} \leq 0.3$ mJy/Beam. Then, the jet-to-counterjet ratio is $J \geq 966.7$ \citep{Ojha2010}. Given the observed radio spectral index of $\alpha = -0.45$ between 2.7 and 5 GHz by \cite{Stickel_1994}, one has in principle all the needed information to efficiently deduce the maximum associated value of the jet misalignment \citep{Urry1995},
\begin{equation}
\theta \leq \arccos \left( \frac{J^{1/(2-\alpha)} - 1}{J^{1/(2-\alpha)} + 1} \, \frac{1}{\beta} \right) \leq 27.6 \, \mathrm{deg},
\end{equation}
assuming the intrinsic flow speed $\beta \rightarrow 1$.

The VHE detection of the source that we report highlights also a significant peculiarity compared to the other VHE radio-galaxies.
The flux and frequency of the synchrotron peak, most likely between UV and X-rays, is also pushing for a BL Lac nature. According to the blazar-envelope unification scheme \citep{Meyer_2011}, this synchrotron peak position is actually the most extreme one among all known VHE radio-galaxies, as presented in Figure 8 in \cite{Fukazawa_2015}.

The redshift of the source ($z = 0.055$) is very high compared to the other VHE radio-galaxies. The farthest VHE radio-galaxies yet observed is NGC 1275 with a redshift of $z = 0.017$, and IC~310 with $z = 0.0189$ when considering misaligned blazars. If strongly misaligned, this would imply an extremely powerful jet.
Also,  \cite{Wills2004} have shown that the nuclei of BL Lac type objects and FR~I radio-galaxies can be used to discriminate the nature following the intensity of their [OIII] emission line, with Log$_{10}$ $L_{[OIII], BL Lac} = 40.840 \pm 0.156$ erg s$^{-1}$  and Log $L_{[OIII], FRI} = 39.509 \pm 0.213$ erg s$^{-1}$. The measured [OIII] line of \pks\ fall within the BL Lac category with a luminosity of  Log$_{10}$ $L_{[OIII]} = 40.64$ erg s$^{-1}$.

Finally, in order to reproduce the observed MWL SED, and especially the VHE spectrum, the leptonic and lepto-hadronic models presented here naturally need significant Doppler boosting of the emission associate to small angles with the line of sight. 
The X-ray variability give a constrains on the maximal possible misalignment of $\theta_{\mathrm{max}} \simeq 15$ deg. Throughout this paper we did not explore the effects of external inverse-Compton emission between different non-thermal zones which may loose the viewing angle constraint. This scenario has been successfully applied to describe wide inverse-Compton emissions from X rays to VHE for radio-galaxies, as M~87 or NGC~1275 \citep{Tavecchio_2008,Tavecchio_2014}. 
Even if the high energy bump observed in the SED does not reveal an extra wideness expected by such a scenario, such a a possibility could be tested by future  MWL observations.

Hence \pks\, is an additional VHE emitting AGN in the blurry zone between being  blazar and radio-galaxy-like in nature.
According to MWL observations and different aspects of the source presented in this paper, both interpretations of \pks\ should be borne in mind. 
Results of modelling could evolve by having further more simultaneous observations.
\section{Summary}
In this paper the discovery of the  VHE $\gamma$-ray emission from \pks\ is reported. The significance of the detection is 6.1$\sigma$ following the exposure of
 5.5 hours. The VHE source is well characterized by a power-law function with a photon index  $\Gamma=2.84\pm0.50$ and normalization  $N_0(1~\mathrm{TeV})=(2.78\pm0.70)\times10^{-12}$\,TeV$^{-1}$cm$^{-2}$s$^{-1}$. 
H.E.S.S. monitoring of the object  is augmented with multi-frequency observations  collected with \textit{Fermi}-LAT, \textit{Swift}-XRT, \textit{Swift}-UVOT, ATOM and WISE.
Pronounced variability is observed at HE and X-ray regimes, while in VHE, optical and IR bands no significant variability has been found. 
However, the past observations of  high UV emission with a hard spectrum  point toward a variability of the nuclear emission.

None of the leptonic or lepto-hadronic models can be rejected by our current MWL data, even if the energetics  favour a leptonic scenario.
Only a further MWL campaign during a high level of source activity 
allow further probe of the nature of the non-thermal emission.

The current classification of \pks\ as a regular radio-galaxy is problematic due to its unusual properties. Therefore, we argue that a classification as a blazar, or moderately misaligned blazar, is strongly favoured. 
However, only more simultaneous further MWL observation campaigns would provide the final word of this peculiar source classification.

\section*{Acknowledgements}
The support of the Namibian authorities and of the University of Namibia in facilitating the construction and operation of H.E.S.S. is gratefully acknowledged, as is the support by the German Ministry for Education and Research (BMBF), the Max Planck Society, the German Research Foundation (DFG), the Alexander von Humboldt Foundation, the Deutsche Forschungsgemeinschaft, the French Ministry for Research, the CNRS-IN2P3 and the Astroparticle Interdisciplinary Programme of the CNRS, the U.K. Science and Technology Facilities Council (STFC), the IPNP of the Charles University, the Czech Science Foundation, the Polish National Science Centre, the South African Department of Science and Technology and National Research Foundation, the University of Namibia, the National Commission on Research, Science \& Technology of Namibia (NCRST), the Innsbruck University, the Austrian Science Fund (FWF), and the Austrian Federal Ministry for Science, Research and Economy, the University of Adelaide and the Australian Research Council, the Japan Society for the Promotion of Science and by the University of Amsterdam.
We appreciate the excellent work of the technical support staff in Berlin, Durham, Hamburg, Heidelberg, Palaiseau, Paris, Saclay, and in Namibia in the construction and operation of the equipment. This work benefited from services provided by the H.E.S.S. Virtual Organisation, supported by the national resource providers of the EGI Federation.
A. Wierzcholska is supported by the Foundation for Polish Science (FNP). O.Hervet thanks the U.S. National Science Foundation for support under grant PHY-1307311 and the Observatoire de Paris for financial support with ATER position.
\bibliographystyle{mn2e_williams}
\bibliography{references}

\begin{thebibliography}{57}
\expandafter\ifx\csname natexlab\endcsname\relax\def\natexlab#1{#1}\fi

\bibitem[{{Abdo} {et~al}\mbox{.}(2010{\natexlab{a}}){Abdo}, {Ackermann},
  {Ajello}, {Allafort}, {Antolini}, {Atwood}, {Axelsson}, {Baldini}, {Ballet},
  {Barbiellini}, \& et~al.}]{1FGL}
{Abdo} A.~A. {et~al.}, 2010{\natexlab{a}}, \apjs, 188, 405

\bibitem[{{Abdo} {et~al}\mbox{.}(2010{\natexlab{b}}){Abdo}, {Ackermann},
  {Ajello}, {Baldini}, {Ballet}, {Barbiellini}, {Bastieri}, {Bechtol},
  {Bellazzini}, {Berenji}, {Blandford}, {Bloom}, {Bonamente}, {Borgland},
  {Bouvier}, {Brandt}, {Bregeon}, {Brez}, {Brigida}, {Bruel}, {Buehler},
  {Burnett}, {Buson}, {Caliandro}, {Cameron}, {Cannon}, {Caraveo}, {Carrigan},
  {Casandjian}, {Cavazzuti}, {Cecchi}, {{\c C}elik}, {Celotti}, {Charles},
  {Chekhtman}, {Chen}, {Cheung}, {Chiang}, {Ciprini}, {Claus}, {Cohen-Tanugi},
  {Colafrancesco}, {Conrad}, {Davis}, {Dermer}, {de Angelis}, {de Palma},
  {Silva}, {Drell}, {Dubois}, {Favuzzi}, {Fegan}, {Ferrara}, {Fortin},
  {Frailis}, {Fukazawa}, {Fusco}, {Gargano}, {Gasparrini}, {Gehrels},
  {Germani}, {Giglietto}, {Giommi}, {Giordano}, {Giroletti}, {Glanzman},
  {Godfrey}, {Grandi}, {Grenier}, {Grove}, {Guillemot}, {Guiriec}, {Hadasch},
  {Hayashida}, {Hays}, {Horan}, {Hughes}, {Jackson}, {J{\'o}hannesson},
  {Johnson}, {Johnson}, {Kamae}, {Katagiri}, {Kataoka}, {Kn{\"o}dlseder},
  {Kuss}, {Lande}, {Latronico}, {Lee}, {Lemoine-Goumard}, {Llena Garde},
  {Longo}, {Loparco}, {Lott}, {Lovellette}, {Lubrano}, {Madejski}, {Makeev},
  {Malaguti}, {Mazziotta}, {McConville}, {McEnery}, {Michelson}, {Migliori},
  {Mitthumsiri}, {Mizuno}, {Monte}, {Monzani}, {Morselli}, {Moskalenko},
  {Murgia}, {Naumann-Godo}, {Nestoras}, {Nolan}, {Norris}, {Nuss}, {Ohsugi},
  {Okumura}, {Omodei}, {Orlando}, {Ormes}, {Paneque}, {Panetta}, {Parent},
  {Pelassa}, {Pepe}, {Persic}, {Pesce-Rollins}, {Piron}, {Porter}, {Rain{\`o}},
  {Rando}, {Razzano}, {Razzaque}, {Reimer}, {Reimer}, {Reyes}, {Roth},
  {Sadrozinski}, {Sanchez}, {Sander}, {Scargle}, {Sgr{\`o}}, {Siskind},
  {Smith}, {Spandre}, {Spinelli}, {Stawarz}, {Stecker}, {Strickman}, {Suson},
  {Takahashi}, {Tanaka}, {Thayer}, {Thayer}, {Thompson}, {Tibaldo}, {Torres},
  {Torresi}, {Tosti}, {Tramacere}, {Uchiyama}, {Usher}, {Vandenbroucke},
  {Vasileiou}, {Vilchez}, {Villata}, {Vitale}, {Waite}, {Wang}, {Winer},
  {Wood}, {Yang}, {Ylinen}, \& {Ziegler}}]{Abdo_2010b}
{Abdo} A.~A. {et~al.}, 2010{\natexlab{b}}, \apj, 720, 912

\bibitem[{{Acero} {et~al}\mbox{.}(2015){Acero}, {Ackermann}, {Ajello},
  {Albert}, {Atwood}, {Axelsson}, {Baldini}, {Ballet}, {Barbiellini},
  {Bastieri}, {Belfiore}, {Bellazzini}, {Bissaldi}, {Blandford}, \& {Fermi-LAT
  Collaboration}}]{3FGL}
{Acero} F. {et~al.}, 2015, \apjs, 218, 23

\bibitem[{{Aharonian} {et~al}\mbox{.}(2006{\natexlab{a}}){Aharonian},
  {Akhperjanian}, {Bazer-Bachi}, {Beilicke}, {Benbow}, {Berge}, {Bernl{\"o}hr},
  {Boisson}, {Bolz}, {Borrel}, {Braun}, {Breitling}, {Brown}, {B{\"u}hler},
  {B{\"u}sching}, {Carrigan}, {Chadwick}, {Chounet}, {Cornils}, {Costamante},
  {Degrange}, {Dickinson}, {Djannati-Ata{\"i}}, {O'C.~Drury}, {Dubus},
  {Egberts}, {Emmanoulopoulos}, {Espigat}, {Feinstein}, {Ferrero}, {Fiasson},
  {Fontaine}, {Funk}, {Funk}, {Gallant}, {Giebels}, {Glicenstein}, {Goret},
  {Hadjichristidis}, {Hauser}, {Hauser}, {Heinzelmann}, {Henri}, {Hermann},
  {Hinton}, {Hofmann}, {Holleran}, {Horns}, {Jacholkowska}, {de Jager},
  {Kh{\'e}lifi}, {Komin}, {Konopelko}, {Kosack}, {Latham}, {Le Gallou},
  {Lemi{\`e}re}, {Lemoine-Goumard}, {Lohse}, {Martin}, {Martineau-Huynh},
  {Marcowith}, {Masterson}, {McComb}, {de Naurois}, {Nedbal}, {Nolan},
  {Noutsos}, {Orford}, {Osborne}, {Ouchrif}, {Panter}, {Pelletier}, {Pita},
  {P{\"u}hlhofer}, {Punch}, {Raubenheimer}, {Raue}, {Rayner}, {Reimer},
  {Reimer}, {Ripken}, {Rob}, {Rolland}, {Rowell}, {Sahakian}, {Saug{\'e}},
  {Schlenker}, {Schlickeiser}, {Schwanke}, {Sol}, {Spangler}, {Spanier},
  {Steenkamp}, {Stegmann}, {Superina}, {Tavernet}, {Terrier}, {Th{\'e}oret},
  {Tluczykont}, {van Eldik}, {Vasileiadis}, {Venter}, {Vincent}, {V{\"o}lk},
  {Wagner}, \& {Ward}}]{Aharonian2006_crab}
{Aharonian} F. {et~al.}, 2006{\natexlab{a}}, \aap, 457, 899

\bibitem[{{Aharonian} {et~al}\mbox{.}(2006{\natexlab{b}}){Aharonian},
  {Akhperjanian}, {Bazer-Bachi}, {Beilicke}, {Benbow}, {Berge}, {Bernl{\"o}hr},
  {Boisson}, {Bolz}, {Borrel}, {Braun}, {Breitling}, {Brown}, {B{\"u}hler},
  {B{\"u}sching}, {Carrigan}, {Chadwick}, {Chounet}, {Cornils}, {Costamante},
  {Degrange}, {Dickinson}, {Djannati-Ata{\"i}}, {O'C.~Drury}, {Dubus},
  {Egberts}, {Emmanoulopoulos}, {Espigat}, {Feinstein}, {Ferrero}, {Fiasson},
  {Fontaine}, {Funk}, {Funk}, {Gallant}, {Giebels}, {Glicenstein}, {Goret},
  {Hadjichristidis}, {Hauser}, {Hauser}, {Heinzelmann}, {Henri}, {Hermann},
  {Hinton}, {Hofmann}, {Holleran}, {Horns}, {Jacholkowska}, {de Jager},
  {Kh{\'e}lifi}, {Komin}, {Konopelko}, {Kosack}, {Latham}, {Le Gallou},
  {Lemi{\`e}re}, {Lemoine-Goumard}, {Lohse}, {Martin}, {Martineau-Huynh},
  {Marcowith}, {Masterson}, {McComb}, {de Naurois}, {Nedbal}, {Nolan},
  {Noutsos}, {Orford}, {Osborne}, {Ouchrif}, {Panter}, {Pelletier}, {Pita},
  {P{\"u}hlhofer}, {Punch}, {Raubenheimer}, {Raue}, {Rayner}, {Reimer},
  {Reimer}, {Ripken}, {Rob}, {Rolland}, {Rowell}, {Sahakian}, {Saug{\'e}},
  {Schlenker}, {Schlickeiser}, {Schwanke}, {Sol}, {Spangler}, {Spanier},
  {Steenkamp}, {Stegmann}, {Superina}, {Tavernet}, {Terrier}, {Th{\'e}oret},
  {Tluczykont}, {van Eldik}, {Vasileiadis}, {Venter}, {Vincent}, {V{\"o}lk},
  {Wagner}, \& {Ward}}]{Crab_paper}
{Aharonian} F. {et~al.}, 2006{\natexlab{b}}, \aap, 457, 899

\bibitem[{{Aharonian}(2000)}]{Aharonian_2000}
{Aharonian} F.~A., 2000, \na, 5, 377

\bibitem[{{Atwood} {et~al}\mbox{.}(2013){Atwood}, {Albert}, {Baldini},
  {Tinivella}, {Bregeon}, {Pesce-Rollins}, {Sgr{\`o}}, {Bruel}, {Charles},
  {Drlica-Wagner}, {Franckowiak}, {Jogler}, {Rochester}, {Usher}, {Wood},
  {Cohen-Tanugi}, \& {S.~Zimmer for the Fermi-LAT
  Collaboration}}]{Atwood2013_pass8}
{Atwood} W. {et~al.}, 2013, ArXiv e-prints

\bibitem[{{Bettoni} {et~al}\mbox{.}(2003){Bettoni}, {Falomo}, {Fasano}, \&
  {Govoni}}]{Bettoni_2003}
{Bettoni} D., {Falomo} R., {Fasano} G., {Govoni} F., 2003, \aap, 399, 869

\bibitem[{{B{\"o}ttcher} {et~al}\mbox{.}(2013){B{\"o}ttcher}, {Reimer},
  {Sweeney}, \& {Prakash}}]{Boettcher_2013}
{B{\"o}ttcher} M., {Reimer} A., {Sweeney} K., {Prakash} A., 2013, \apj, 768, 54

\bibitem[{{Burrows} {et~al}\mbox{.}(2005){Burrows}, {Hill}, {Nousek}, {Kennea},
  {Wells}, {Osborne}, {Abbey}, {Beardmore}, {Mukerjee}, {Short}, {Chincarini},
  {Campana}, {Citterio}, {Moretti}, {Pagani}, {Tagliaferri}, {Giommi},
  {Capalbi}, {Tamburelli}, {Angelini}, {Cusumano}, {Br{\"a}uninger}, {Burkert},
  \& {Hartner}}]{Burrows20015}
{Burrows} D.~N. {et~al.}, 2005, \ssr, 120, 165

\bibitem[{{Cerruti} {et~al}\mbox{.}(2015){Cerruti}, {Zech}, {Boisson}, \&
  {Inoue}}]{Cerruti_2015}
{Cerruti} M., {Zech} A., {Boisson} C., {Inoue} S., 2015, \mnras, 448, 910

\bibitem[{{Condon} {et~al}\mbox{.}(1998){Condon}, {Cotton}, {Greisen}, {Yin},
  {Perley}, {Taylor}, \& {Broderick}}]{Condon_1998}
{Condon} J.~J., {Cotton} W.~D., {Greisen} E.~W., {Yin} Q.~F., {Perley} R.~A.,
  {Taylor} G.~B., {Broderick} J.~J., 1998, \aj, 115, 1693

\bibitem[{{de Naurois} \& {Rolland}(2009)}]{Naurois2009}
{de Naurois} M., {Rolland} L., 2009, Astroparticle Physics, 32, 231

\bibitem[{{Ekers} {et~al}\mbox{.}(1989){Ekers}, {Wall}, {Shaver}, {Goss},
  {Fosbury}, {Danziger}, {Moorwood}, {Malin}, {Monk}, \& {Ekers}}]{Ekers1989}
{Ekers} R.~D. {et~al.}, 1989, \mnras, 236, 737

\bibitem[{{Feldman} \& {Cousins}(1998)}]{Feldman98}
{Feldman} G.~J., {Cousins} R.~D., 1998, \prd, 57, 3873

\bibitem[{{Finke} {et~al}\mbox{.}(2010){Finke}, {Razzaque}, \&
  {Dermer}}]{Finke_2010}
{Finke} J.~D., {Razzaque} S., {Dermer} C.~D., 2010, \apj, 712, 238

\bibitem[{{Fioc} \& {Rocca-Volmerange}(1999)}]{Fioc_1999}
{Fioc} M., {Rocca-Volmerange} B., 1999, eprint arXiv:astro-ph/9912179

\bibitem[{{Fomalont} {et~al}\mbox{.}(2000){Fomalont}, {Frey}, {Paragi},
  {Gurvits}, {Scott}, {Taylor}, {Edwards}, \& {Hirabayashi}}]{Fomalont2000}
{Fomalont} E.~B., {Frey} S., {Paragi} Z., {Gurvits} L.~I., {Scott} W.~K.,
  {Taylor} A.~R., {Edwards} P.~G., {Hirabayashi} H., 2000, \apjs, 131, 95

\bibitem[{{Fukazawa} {et~al}\mbox{.}(2015){Fukazawa}, {Finke}, {Stawarz},
  {Tanaka}, {Itoh}, \& {Tokuda}}]{Fukazawa_2015}
{Fukazawa} Y., {Finke} J., {Stawarz} {\L}., {Tanaka} Y., {Itoh} R., {Tokuda}
  S., 2015, \apj, 798, 74

\bibitem[{{Giommi} {et~al}\mbox{.}(2006){Giommi}, {Blustin}, {Capalbi},
  {Colafrancesco}, {Cucchiara}, {Fuhrmann}, {Krimm}, {Marchili}, {Massaro},
  {Perri}, {Tagliaferri}, {Tosti}, {Tramacere}, {Burrows}, {Chincarini},
  {Falcone}, {Gehrels}, {Kennea}, \& {Sambruna}}]{Giommi2006}
{Giommi} P. {et~al.}, 2006, \aap, 456, 911

\bibitem[{{Giovannini} {et~al}\mbox{.}(1994){Giovannini}, {Feretti}, {Venturi},
  {Lara}, {Marcaide}, {Rioja}, {Spangler}, \& {Wehrle}}]{Giovannini_1994}
{Giovannini} G., {Feretti} L., {Venturi} T., {Lara} L., {Marcaide} J., {Rioja}
  M., {Spangler} S.~R., {Wehrle} A.~E., 1994, \apj, 435, 116

\bibitem[{{Gliozzi} {et~al}\mbox{.}(2008){Gliozzi}, {Foschini}, {Sambruna}, \&
  {Tavecchio}}]{Gliozzi_2008}
{Gliozzi} M., {Foschini} L., {Sambruna} R.~M., {Tavecchio} F., 2008, \aap, 478,
  723

\bibitem[{{Hauser} {et~al}\mbox{.}(2004){Hauser}, {M{\"o}llenhoff},
  {P{\"u}hlhofer}, {Wagner}, {Hagen}, \& {Knoll}}]{Hauser2004}
{Hauser} M., {M{\"o}llenhoff} C., {P{\"u}hlhofer} G., {Wagner} S.~J., {Hagen}
  H.-J., {Knoll} M., 2004, Astronomische Nachrichten, 325, 659

\bibitem[{{Hervet} {et~al}\mbox{.}(2015){Hervet}, {Boisson}, \&
  {Sol}}]{Hervet_2015}
{Hervet} O., {Boisson} C., {Sol} H., 2015, \aap, 578, A69

\bibitem[{{H.E.S.S.~Collaboration}
  {et~al}\mbox{.}(2015){H.E.S.S.~Collaboration}, {Abramowski}, {Aharonian},
  {Ait Benkhali}, {Akhperjanian}, {Ang{\"u}ner}, {Anton}, {Backes},
  {Balenderan}, {Balzer}, \& et~al.}]{hess_ap}
{H.E.S.S.~Collaboration} {et~al.}, 2015, \aap, 573, A31

\bibitem[{{Inskip} {et~al}\mbox{.}(2010){Inskip}, {Tadhunter}, {Morganti},
  {Holt}, {Ramos Almeida}, \& {Dicken}}]{Inskip_2010}
{Inskip} K.~J., {Tadhunter} C.~N., {Morganti} R., {Holt} J., {Ramos Almeida}
  C., {Dicken} D., 2010, \mnras, 407, 1739

\bibitem[{{Jones} {et~al}\mbox{.}(2009){Jones}, {Read}, {Saunders}, {Colless},
  {Jarrett}, {Parker}, {Fairall}, {Mauch}, {Sadler}, {Watson}, {Burton},
  {Campbell}, {Cass}, {Croom}, {Dawe}, {Fiegert}, {Frankcombe}, {Hartley},
  {Huchra}, {James}, {Kirby}, {Lahav}, {Lucey}, {Mamon}, {Moore}, {Peterson},
  {Prior}, {Proust}, {Russell}, {Safouris}, {Wakamatsu}, {Westra}, \&
  {Williams}}]{Jones2009}
{Jones} D.~H. {et~al.}, 2009, \mnras, 399, 683

\bibitem[{{Kalberla} {et~al}\mbox{.}(2005){Kalberla}, {Burton}, {Hartmann},
  {Arnal}, {Bajaja}, {Morras}, \& {P{\"o}ppel}}]{Kalberla05}
{Kalberla} P.~M.~W., {Burton} W.~B., {Hartmann} D., {Arnal} E.~M., {Bajaja} E.,
  {Morras} R., {P{\"o}ppel} W.~G.~L., 2005, \aap, 440, 775

\bibitem[{{Katarzy{\'n}ski} {et~al}\mbox{.}(2001){Katarzy{\'n}ski}, {Sol}, \&
  {Kus}}]{Katarzynski_2001}
{Katarzy{\'n}ski} K., {Sol} H., {Kus} A., 2001, \aap, 367, 809

\bibitem[{{Li} \& {Ma}(1983)}]{Li1983}
{Li} T.-P., {Ma} Y.-Q., 1983, \apj, 272, 317

\bibitem[{{Mastichiadis} {et~al}\mbox{.}(2013){Mastichiadis}, {Petropoulou}, \&
  {Dimitrakoudis}}]{Mastichiadis_2013}
{Mastichiadis} A., {Petropoulou} M., {Dimitrakoudis} S., 2013, \mnras, 434,
  2684

\bibitem[{{Mattox} {et~al}\mbox{.}(1996){Mattox}, {Bertsch}, {Chiang},
  {Dingus}, {Digel}, {Esposito}, {Fierro}, {Hartman}, {Hunter}, {Kanbach},
  {Kniffen}, {Lin}, {Macomb}, {Mayer-Hasselwander}, {Michelson}, {von
  Montigny}, {Mukherjee}, {Nolan}, {Ramanamurthy}, {Schneid}, {Sreekumar},
  {Thompson}, \& {Willis}}]{Mattox96}
{Mattox} J.~R. {et~al.}, 1996, \apj, 461, 396

\bibitem[{{Mauch} {et~al}\mbox{.}(2003){Mauch}, {Murphy}, {Buttery}, {Curran},
  {Hunstead}, {Piestrzynski}, {Robertson}, \& {Sadler}}]{Mauch_2003}
{Mauch} T., {Murphy} T., {Buttery} H.~J., {Curran} J., {Hunstead} R.~W.,
  {Piestrzynski} B., {Robertson} J.~G., {Sadler} E.~M., 2003, \mnras, 342, 1117

\bibitem[{{Meyer} {et~al}\mbox{.}(2011){Meyer}, {Fossati}, {Georganopoulos}, \&
  {Lister}}]{Meyer_2011}
{Meyer} E.~T., {Fossati} G., {Georganopoulos} M., {Lister} M.~L., 2011, \apj,
  740, 98

\bibitem[{{M{\"u}cke} \& {Protheroe}(2000)}]{Mucke_2000}
{M{\"u}cke} A., {Protheroe} R.~J., 2000, in American Institute of Physics
  Conference Series, Vol. 515, American Institute of Physics Conference Series,
  {Dingus} B.~L., {Salamon} M.~H., {Kieda} D.~B., eds., pp. 149--153

\bibitem[{{M{\"u}ller} {et~al}\mbox{.}(2013){M{\"u}ller}, {Krauss}, {Kadler},
  {Tr{\"u}stedt}, {Ojha}, {Ros}, {Wilms}, {B{\"o}ck}, {Dutka}, {Carpenter}, \&
  {the TANAMI collaboration}}]{Muller_2013}
{M{\"u}ller} C. {et~al.}, 2013, ArXiv e-prints: 1301.4384

\bibitem[{{Murphy} {et~al}\mbox{.}(2010){Murphy}, {Sadler}, {Ekers},
  {Massardi}, {Hancock}, {Mahony}, {Ricci}, {Burke-Spolaor}, {Calabretta},
  {Chhetri}, {de Zotti}, {Edwards}, {Ekers}, {Jackson}, {Kesteven}, {Lindley},
  {Newton-McGee}, {Phillips}, {Roberts}, {Sault}, {Staveley-Smith},
  {Subrahmanyan}, {Walker}, \& {Wilson}}]{Murphy_2010}
{Murphy} T. {et~al.}, 2010, \mnras, 402, 2403

\bibitem[{{Nesci} {et~al}\mbox{.}(2013){Nesci}, {Tosti}, {Pursimo}, {Ojha}, \&
  {Kadler}}]{Nesci_2013}
{Nesci} R., {Tosti} G., {Pursimo} T., {Ojha} R., {Kadler} M., 2013, \aap, 555,
  A2

\bibitem[{{Nolan} {et~al}\mbox{.}(2012){Nolan}, {Abdo}, {Ackermann}, {Ajello},
  {Allafort}, {Antolini}, {Atwood}, {Axelsson}, {Baldini}, {Ballet}, \&
  et~al.}]{2FGL}
{Nolan} P.~L. {et~al.}, 2012, \apjs, 199, 31

\bibitem[{{Ojha} {et~al}\mbox{.}(2010){Ojha}, {Kadler}, {B{\"o}ck}, {Booth},
  {Dutka}, {Edwards}, {Fey}, {Fuhrmann}, {Gaume}, {Hase}, {Horiuchi},
  {Jauncey}, {Johnston}, {Katz}, {Lister}, {Lovell}, {M{\"u}ller}, {Pl{\"o}tz},
  {Quick}, {Ros}, {Taylor}, {Thompson}, {Tingay}, {Tosti}, {Tzioumis}, {Wilms},
  \& {Zensus}}]{Ojha2010}
{Ojha} R. {et~al.}, 2010, \aap, 519, A45

\bibitem[{{Parsons} \& {Hinton}(2014)}]{Parsons2014}
{Parsons} R.~D., {Hinton} J.~A., 2014, Astroparticle Physics, 56, 26

\bibitem[{{Piron} {et~al}\mbox{.}(2001){Piron}, {Djannati-Atai}, {Punch},
  {Tavernet}, {Barrau}, {Bazer-Bachi}, {Chounet}, {Debiais}, {Degrange},
  {Dezalay}, {Espigat}, {Fabre}, {Fleury}, {Fontaine}, {Goret}, {Gouiffes},
  {Khelifi}, {Malet}, {Masterson}, {Mohanty}, {Nuss}, {Renault}, {Rivoal},
  {Rob}, \& {Vorobiov}}]{Piron2001}
{Piron} F. {et~al.}, 2001, \aap, 374, 895

\bibitem[{{Poole} {et~al}\mbox{.}(2008){Poole}, {Breeveld}, {Page}, {Landsman},
  {Holland}, {Roming}, {Kuin}, {Brown}, {Gronwall}, {Hunsberger}, {Koch},
  {Mason}, {Schady}, {vanden Berk}, {Blustin}, {Boyd}, {Broos}, {Carter},
  {Chester}, {Cucchiara}, {Hancock}, {Huckle}, {Immler}, {Ivanushkina},
  {Kennedy}, {Marshall}, {Morgan}, {Pandey}, {de Pasquale}, {Smith}, \&
  {Still}}]{Poole08}
{Poole} T.~S. {et~al.}, 2008, \mnras, 383, 627

\bibitem[{{Potter} \& {Cotter}(2012)}]{Potter_2012}
{Potter} W.~J., {Cotter} G., 2012, \mnras, 423, 756

\bibitem[{{Protheroe} \& {M{\"u}cke}(2001)}]{Protheroe_2001}
{Protheroe} R.~J., {M{\"u}cke} A., 2001, in Astronomical Society of the Pacific
  Conference Series, Vol. 250, Particles and Fields in Radio Galaxies
  Conference, {Laing} R.~A., {Blundell} K.~M., eds., p. 113

\bibitem[{{Reines} \& {Volonteri}(2015)}]{Reines_2015}
{Reines} A.~E., {Volonteri} M., 2015, \apj, 813, 82

\bibitem[{{Roming} {et~al}\mbox{.}(2005){Roming}, {Kennedy}, {Mason}, {Nousek},
  {Ahr}, {Bingham}, {Broos}, {Carter}, {Hancock}, {Huckle}, {Hunsberger},
  {Kawakami}, {Killough}, {Koch}, {McLelland}, {Smith}, {Smith}, {Soto},
  {Boyd}, {Breeveld}, {Holland}, {Ivanushkina}, {Pryzby}, {Still}, \&
  {Stock}}]{Roming2005}
{Roming} P.~W.~A. {et~al.}, 2005, \ssr, 120, 95

\bibitem[{{Schlafly} \& {Finkbeiner}(2011)}]{Schlafly}
{Schlafly} E.~F., {Finkbeiner} D.~P., 2011, \apj, 737, 103

\bibitem[{{Stickel} {et~al}\mbox{.}(1994){Stickel}, {Meisenheimer}, \&
  {Kuehr}}]{Stickel_1994}
{Stickel} M., {Meisenheimer} K., {Kuehr} H., 1994, \aaps, 105

\bibitem[{{Tavecchio} \& {Ghisellini}(2008)}]{Tavecchio_2008}
{Tavecchio} F., {Ghisellini} G., 2008, \mnras, 385, L98

\bibitem[{{Tavecchio} \& {Ghisellini}(2014)}]{Tavecchio_2014}
{Tavecchio} F., {Ghisellini} G., 2014, \mnras, 443, 1224

\bibitem[{{Trussoni} {et~al}\mbox{.}(1999){Trussoni}, {Vagnetti}, {Massaglia},
  {Feretti}, {Parma}, {Morganti}, {Fanti}, \& {Padovani}}]{Trussoni99}
{Trussoni} E., {Vagnetti} F., {Massaglia} S., {Feretti} L., {Parma} P.,
  {Morganti} R., {Fanti} R., {Padovani} P., 1999, \aap, 348, 437

\bibitem[{{Urry} \& {Padovani}(1995)}]{Urry1995}
{Urry} C.~M., {Padovani} P., 1995, \pasp, 107, 803

\bibitem[{{Vaughan} {et~al}\mbox{.}(2003){Vaughan}, {Edelson}, {Warwick}, \&
  {Uttley}}]{Vaughan2003}
{Vaughan} S., {Edelson} R., {Warwick} R.~S., {Uttley} P., 2003, \mnras, 345,
  1271

\bibitem[{{Venturi} {et~al}\mbox{.}(2000){Venturi}, {Morganti}, {Tzioumis}, \&
  {Reynolds}}]{Venturi2000}
{Venturi} T., {Morganti} R., {Tzioumis} T., {Reynolds} J., 2000, \aap, 363, 84

\bibitem[{{Wills} {et~al}\mbox{.}(2004){Wills}, {Morganti}, {Tadhunter},
  {Robinson}, \& {Villar-Martin}}]{Wills2004}
{Wills} K.~A., {Morganti} R., {Tadhunter} C.~N., {Robinson} T.~G.,
  {Villar-Martin} M., 2004, \mnras, 347, 771

\bibitem[{{Wright} {et~al}\mbox{.}(2010){Wright}, {Eisenhardt}, {Mainzer},
  {Ressler}, {Cutri}, {Jarrett}, {Kirkpatrick}, {Padgett}, {McMillan},
  {Skrutskie}, {Stanford}, {Cohen}, {Walker}, {Mather}, {Leisawitz}, {Gautier},
  {McLean}, {Benford}, {Lonsdale}, {Blain}, {Mendez}, {Irace}, {Duval}, {Liu},
  {Royer}, {Heinrichsen}, {Howard}, {Shannon}, {Kendall}, {Walsh}, {Larsen},
  {Cardon}, {Schick}, {Schwalm}, {Abid}, {Fabinsky}, {Naes}, \&
  {Tsai}}]{Wright_2010}
{Wright} E.~L. {et~al.}, 2010, \aj, 140, 1868

\end{thebibliography}

\affiliations
\label{lastpage}
\end{document}